\documentclass[lettersize,journal]{IEEEtran}
\IEEEoverridecommandlockouts
\usepackage{url}
\usepackage{ifthen}
\usepackage{cite}
\usepackage{amsmath} 
\usepackage{amssymb}
\usepackage{graphicx}
\usepackage{epstopdf}
\usepackage{epsfig}
\usepackage{multirow}
\usepackage{amsthm}
\usepackage{comment}
\usepackage{caption}
\usepackage{enumitem}
\usepackage{color}
\usepackage{bbm}
\usepackage{algorithm}
\usepackage{algpseudocode}
\usepackage{csquotes}
\usepackage{subfiles}
\usepackage{caption, multirow, makecell}
\usepackage{array}
\usepackage{caption}
\usepackage{afterpage}
\usepackage{dblfloatfix}
\usepackage{hyperref}
\usepackage{graphicx}
\usepackage{balance}
\usepackage{subcaption}
\usepackage{diagbox}
\usepackage{float}
\newtheorem{claim}{Claim}
\newtheorem{thm}{Theorem}
\newtheorem{lem}{Lemma}

\newtheorem{cor}{Corollary}
\newtheorem{example}{Example}

\newtheorem{defn}{Definition}

\newtheorem{rem}{Remark}

\def\BibTeX{{\rm B\kern-.05em{\sc i\kern-.025em b}\kern-.08em
		T\kern-.1667em\lower.7ex\hbox{E}\kern-.125emX}}
\begin{document}	
		\title{D2D Coded Caching Schemes for Multiaccess Networks with Combinatorial Access Topology\\}
	
	\author{\IEEEauthorblockN{Rashid Ummer N.T. and B. Sundar Rajan}\\
		\IEEEauthorblockA{Department of Electrical Communication Engineering, Indian Institute of Science, Bangalore, India \\
			E-mail: \{rashidummer, bsrajan\}@iisc.ac.in}
		}
	\maketitle
	
	\begin{abstract}
Device-to-device (D2D) communication is one of the most promising techniques for next-generation wireless Internet of Things networks. This paper considers coded caching in a wireless multiaccess D2D network, where users communicate with each other and can access multiple cache nodes. Access topologies derived from two combinatorial designs known as the $t$-design and $t$-group divisible design ($t$-GDD), referred to as the $t$-design and $t$-GDD topologies, respectively, have been studied recently for the multiaccess coded caching (MACC) network. These access topologies, which subsume the other known topologies except the cyclic wrap-around topology, are extended to a multiaccess D2D coded caching (MADCC) network. Novel MADCC schemes are proposed, and schemes are also derived from existing MACC schemes with $t$-design and $t$-GDD topologies. To compare different MADCC schemes, the metrics of load per user and subpacketization level are used while keeping the number of caches, cache size, and access degree the same. A comparison of the proposed schemes with those derived from existing MACC schemes, as well as with the existing MADCC scheme under a cyclic wrap-around topology, shows that the proposed schemes have an advantage in either load per user or subpacketization level or both. Additionally, several low subpacketization level coded caching schemes are obtained for the original D2D network with dedicated caches.
	\end{abstract}
	\begin{IEEEkeywords}
		D2D coded caching, placement delivery array, multiaccess networks, $t$-design, $t$-group divisible design.
	\end{IEEEkeywords}		
	
	\section{Introduction}
	
	Device-to-device (D2D) communication is one of the most promising techniques for next-generation wireless Internet of Things (IoT) networks \cite{BZ, MACF}. D2D communication also serves as a key enabler for the realization of IoT in future fifth-generation and beyond wireless cellular communication systems \cite{TUY, LLLW}. Many emerging IoT applications require very low latency, and edge caching in IoT networks has been identified as a promising solution to decrease latency \cite{ZTS}. {\footnotetext{ A part of the content has been presented at the IEEE ISIT 2025 held in Ann Arbor (Michigan), USA, June 2025. }} Studying cache-aided D2D networks is therefore of great relevance to IoT networks. In \cite{LZC}, a cache-enabled mobile multi-D2D network has been studied. Motivated by the coded caching technique introduced by Maddah-Ali and Niesen in \cite{MaN} for a broadcast network (referred to as the MAN scheme), coded caching in a wireless D2D network was first studied by Ji, Caire, and Molisch in \cite{Ji} (referred to as the JCM scheme). A D2D network consists of $K$ users, each equipped with a cache of size $M$ files, and a central server with a library of $N$ files. 
	
	A D2D coded caching operates in two phases. First, in the \textit{placement phase}, the central server places data in all user caches. Then, in the \textit{delivery phase}, where the central server is not present, all user demands are served through inter-user coded multicast transmissions. The transmission load of the JCM scheme is shown to be order optimal in \cite{Ji} and \cite{CKRG}. The JCM scheme requires dividing files into $F$ packets (referred to as the subpacketization level), which grows exponentially with the number of users for a given memory ratio $\frac{M}{N}$. Motivated by the concept of Placement Delivery Array (PDA) introduced by Yan \textit{et al.} in \cite{YCT}, Wang \textit{et al.} in \cite{JMQX} proposed an array called D2D Placement Delivery Array (DPDA) which characterizes both the phases of a D2D coded caching scheme with  uncoded placement and one-shot delivery. A procedure to obtain a DPDA from a given regular PDA has been discussed in \cite{JMQX}. Direct construction of a few classes of DPDAs is discussed in \cite{JY} and \cite{DPDA_CRD}.	
	
	Caching at the wireless edge nodes is a widely used technique in communication systems and IoT applications. Motivated by that, a multiaccess coded caching (MACC) model, consisting of a central server with a library of $N$ files connected to $K$ cache-less users and $K$ cache nodes each having a memory of $M$ files, was first proposed in \cite{HKD}. Each user can access, without any access cost, $L$ neighboring cache nodes with a cyclic wrap-around topology. MACC schemes with cyclic wrap-around topology were also studied in \cite{KN,CWLZC,SB}. A MACC scheme using the MAN scheme was proposed in \cite{PDB}. Another access topology using a combinatorial structure called cross resolvable design was considered in \cite{DPB,PDB2,PB}. Cheng \textit{et al.} in \cite{MACC_des} considered the MACC problem with access topology described by two classical combinatorial designs, known as the $t$-design and $t$-group divisible design ($t$-GDD). It was shown in \cite{MACC_des} that the MACC scheme in \cite{PDB} is a special case of the scheme in \cite{MACC_des} using the $t$-design topology, and the access topology using the cross resolvable design considered in \cite{DPB,PDB2,PB} is a special case of the $t$-GDD topology.   
	
	D2D networks in which each user accesses multiple cache nodes have been studied recently for the cyclic wrap-around topology in \cite{WCCWC}, referred to as the \textit{WCCWC (Wu, Cheng, Chen, Wu, Chen) scheme}. A multiaccess D2D coded caching (MADCC) network with other access topologies has not been discussed in the literature. Since the $t$-design and $t$-GDD access topologies subsume the other known network topologies except for the cyclic wrap-around topology, this paper studies the MADCC network under these topologies. As in the existing multiaccess coded caching literature, we also assume that the cache nodes can be accessed without cost. This assumption can be justified by the fact that the access to the cache-nodes can be offloaded on a different network (e.g., a WiFi offloading for hotspots) \cite{CWLZC}. 
	
	In any MADCC problem, the design parameters are the number of users $K$, the number of caches $\Gamma$, the number of files $N$, the cache memory $M$ (in files), and the number of caches a user has access to, $L$ (referred to as the access degree). For a given $\Gamma$, $N$, $M$ and $L$, the number of users $K$ in two different MADCC schemes differs since $K$ depends on the user-to-cache access topology. Therefore, to compare two MADCC schemes with different access topologies, the appropriate metric is the \textit{load per user}, defined as the transmission load $R$ normalized with the number of users $K$. The lower the load per user for a given $\Gamma$, $N$, $M$ and $L$, the better the scheme. Load per user was considered as a metric to compare two MACC schemes, or a MACC scheme and a coded caching scheme with dedicated caches, in \cite{PDB}, \cite{DPB} and \cite{KB}. In this paper, we use the metrics of load per user and subpacketization level to compare different MADCC schemes, while keeping the number of caches, cache size, and access degree the same. 
	
	\subsection{Contributions} 
	We study the wireless D2D coded caching in a multiaccess network with access topology described by a class of combinatorial designs, which has not been previously explored in the literature. The contributions of this paper are summarized below: \\
		\noindent $\bullet$ By using the properties of combinatorial $t$-designs, a novel MADCC scheme with $t$-design topology is proposed. Another MADCC scheme is derived from the MACC scheme with $t$-design topology described in \cite{MACC_des}, referred to as the \textit{derived CWEC (Cheng, Wan, Elia,Caire) scheme with $t$-design topology}. The proposed MADCC scheme with $t$-design topology performs better in terms of subpacketization level while achieving the same load per user compared to the derived CWEC scheme with $t$-design topology. Compared to the WCCWC scheme, the proposed scheme performs better in terms of load per user, while the subpacketization level is equal in some cases and varies (smaller or larger) in others. \\
		\noindent $\bullet$  By using the properties of combinatorial $t$-GDDs and orthogonal arrays, a novel MADCC scheme with $t$-GDD topology is proposed. Another MADCC scheme is derived from the MACC scheme with $t$-GDD topology described in \cite{MACC_des}, referred to as the \textit{derived CWEC scheme with $t$-GDD topology}. The proposed MADCC scheme with $t$-GDD topology performs better in terms of load per user while achieving the same subpacketization level compared to the derived CWEC scheme with $t$-GDD topology in some cases. In other cases, the proposed scheme requires a lesser subpacketization level at the expense of an increase in load per user. Compared to the WCCWC scheme, the proposed scheme with $t$-GDD topology performs better in terms of subpacketization level at the expense of an increase in load per user. \\
		\noindent $\bullet$ Additionally, from the proposed MADCC schemes, several novel classes of low subpacketization level coded caching schemes for the original D2D network with dedicated caches are obtained. In addition to that, another novel D2D scheme for the original D2D network is obtained using $t$-GDDs and orthogonal arrays.
	\subsection{Organization}
	The rest of the paper is organized as follows. Section \ref{prelim_d2d} briefly describes the original D2D network model and reviews the preliminaries of DPDA. Section \ref{prelim_madcc} discusses the MADCC network model and the arrays used to obtain MADCC schemes. Useful definitions and properties of combinatorial designs are presented in Section \ref{designs}. The proposed MADCC scheme with $t$-design topology and its performance analysis are discussed in Section \ref{madcc_schemes_tdes}. The proposed MADCC scheme with $t$-GDD topology and its performance analysis are discussed in Section \ref{madcc_schemes_tgdd}. Coded caching schemes for the original D2D network, derived from the proposed MADCC schemes and a new construction, are discussed in Section \ref{d2d_madcc}. Section \ref{concl_madcc} concludes the paper by discussing the limitations of the proposed schemes and outlining directions for future research.
	
	\textit{Notations}: For any set $\mathcal{A}$, $|\mathcal{A}|$ denotes the cardinality of $\mathcal{A}$. 
	For a set $\mathcal{A}$ and a positive integer $i \leq |\mathcal{A}|$,  $\binom{\mathcal{A}}{i}$ denotes all the $i$-sized subsets of $\mathcal{A}$. For sets $\mathcal{A} \text{ and }\mathcal{B}$, $\mathcal{A} \backslash \mathcal{B}$ denotes the elements in $\mathcal{A}$ but not in $\mathcal{B}$. For any positive integer $n$, $[n]$ denotes the set $\{1,2,...,n\}$. For a vector $\vec{x}$ with $n$ entries and for any $\mathcal{T} \subseteq [n]$, $\vec{x}(\mathcal{T})$ denotes the vector containing the entries of $\vec{x}$ at the indices specified in $\mathcal{T}$. For any two vectors $\vec{x}$ and $\vec{y}$ with same number of entries, $d(\vec{x},\vec{y})$ denotes the number of entries in which $\vec{x}$ and $\vec{y}$ differ. For notational simplicity, we write a vector $\vec{x}=(x_1,x_2,...,x_n)$ in the form $x_1x_2...x_n$. The \textit{occurrence number} of an element $x$ in a sequence (or a collection, such as a set of rows in an array) at position $i$ is defined as the number of times $x$ has appeared in the sequence up to and including position $i$.	
	\section{Original D2D coded caching}\label{prelim_d2d}
	In this section, first we describe the D2D coded caching network model in \cite{Ji} and then review the concept of DPDA proposed in \cite{JMQX} to obtain D2D coded caching schemes. 
	\subsection{Original D2D Network model}
	An $F$-division $(K,M,N)$ D2D coded caching network consists of a central server with a library of $N$ files $\{W_n: n \in [N]\}$ each of size $B$ bits, and $K \le N$ users $\{U_k: k \in [K]\}$, each having a dedicated cache of size $M$ files, where $M<N$. This network is referred to as the \textit{original D2D network} in this paper. The condition $M\geq \frac{N}{K}$ is assumed to ensure that any possible demands can be met using the user cache contents. 

	The D2D coded caching scheme operates in two phases. \\
		\noindent $\bullet$ \textbf{Placement phase}: During this phase, each file $W_{n}$ is split into $F$ non-overlapping packets, i.e., $W_n =\{W_{n,f}: f \in [F]\}$. Each user $U_k$ caches $MB$ bits in its cache denoted by $\mathcal{Z}_k$, with \textit{uncoded data placement}, meaning that the $MB$ bits are taken directly out of the total $NB$ bits in the library. During the placement phase, the later demands of the users are unknown.  \\
		\noindent $\bullet$ \textbf{Delivery phase}:  During this phase, each user requests a file from $\{W_n: n \in [N]\}$, with the set of requested files represented by a demand vector $\vec{d}=(d_1,\ldots,d_K)$. For a given $\vec{d}$, each user broadcasts a coded message $X_{k,\vec{d}}$ consisting of $S_{k,\vec{d}}$ packets, utilizing its cache content and transmitting to all other users over an error-free shared medium. The corresponding worst-case transmission load $R$ normalized to the file size is given by $ R \triangleq \max_{\vec{d}} \frac{\sum_{k=1}^{K}S_{k,\vec{d}}}{F} $.

	\subsection{DPDA}
In this subsection, we review the definition of DPDA and the algorithm for obtaining a D2D-coded caching scheme from a given DPDA. A DPDA is a single array that characterizes both the placement and delivery phases of a D2D coded caching scheme. As a result, constructing a suitable DPDA directly specifies a new D2D coded caching scheme. Low subpacketization level D2D coded caching schemes can be obtained by constructing appropriate DPDAs. A DPDA is a PDA that satisfies an additional property. While a DPDA describes schemes for D2D coded caching networks, a PDA plays an analogous role in centralized shared-link coded caching networks. Hence, we first review the definition of a PDA. 
	\begin{defn}[PDA]\cite{YCT}
		 For positive integers $K, F, Z$ and $S$, an $F \times K$ array $\mathbf{P}=(p_{j,k})$, $j \in [F]$ and $k \in [K]$, composed of a specific symbol $\star$ and $S$ non-negative integers $[S]$, is called a $(K, F, Z, S)$ PDA if it satisfies the following conditions: \\
		\textit{C1}. The symbol $\star$ appears $Z$ times in each column.\\
		\textit{C2}. Each integer occurs at least once in the array.\\
		\textit{C3}. For any two distinct entries $p_{j_1,k_1}$ and $p_{j_2,k_2}$, $p_{j_1,k_1}=p_{j_2,k_2}=s$ is an integer only if (\textit{a}) $j_1 \neq j_2$, $k_1 \neq k_2$, i.e., they lie in distinct rows and distinct columns, and (\textit{b}) $p_{j_1,k_2}=p_{j_2,k_1}=\star$.
	\end{defn} 
	If each integer appears $g$ times in $\mathbf{P}$ where $g$ is a constant, then $\mathbf{P}$ is said to be a $g$-regular $(K, F, Z, S)$ PDA. Given a PDA, a coded caching scheme can be obtained for a shared-link coded caching network \cite{YCT}. Next, we review the definition of a DPDA.  
	\begin{defn}[DPDA]\cite{JMQX}
		A $(K, F, Z, S)$ PDA $\mathbf{P}=(p_{j,k})$, $j \in [F]$ and $k \in [K]$ is called a $(K, F, Z, S)$ DPDA if an additional property described below is satisfied. \\
		\textit{C4}. There is a mapping $\phi$ from $[S]$ to the columns $[K]$ such that if $p_{j,k}=s$ for any $s \in [S]$, then $p_{j,\phi(s)}=\star$.
	\end{defn} 
	In a $(K, F, Z, S)$ DPDA, a column $k \in [K]$ represents a user $U_k$, and a row $j \in [F]$ represents a packet $W_{n,j}$ of each file $W_n$. An entry $p_{j,k}=\star$ indicates that, in the placement phase, the $j^{th}$ packet of all the files is placed in the cache of user $U_k$. Therefore, each user $U_k$ need those packets $W_{d_k,j}$ of its demanded file $W_{d_k}$ such that $p_{j,k}=s$. Corresponding to each $s \in [S]$, the user $U_{\phi(s)}$ transmits a coded multicast message, which is the XOR of the demanded file packets indicated by $s$. Condition $\textit{C3}$ of the DPDA definition ensures that, from a message useful to a user, the user can obtain a required packet of the demanded file, since all other packets in that multicast message are already available in its cache. Condition $\textit{C4}$ of the DPDA definition ensures that, corresponding to each $s \in [S]$, there exists a user $U_{\phi(s)}$ which can transmit the corresponding multicast message. Thus, using Algorithm \ref{d2d_alg}, one can obtain a D2D coded caching scheme from a given DPDA, as proved in \cite{JMQX}, which is stated in the following Theorem. 
	
	\begin{thm}\label{thm:d2d}\cite{JMQX}
		For a given $(K, F, Z, S)$ DPDA $\mathbf{P}=(p_{j,k})_{F \times K}$, a $(K,M,N)$ D2D coded caching scheme can be obtained with subpacketization $F$ and $\frac{M}{N}=\frac{Z}{F}$ using Algorithm \ref{d2d_alg}. For any demand vector $\vec{d}$, the demands of all the users are met with a transmission load of $R=\frac{S}{F}$.
	\end{thm}
	\begin{algorithm}[h]
		\renewcommand{\thealgorithm}{1}
		\caption{\cite{JMQX} D2D Coded caching scheme based on DPDA }
		\label{d2d_alg}
		\begin{algorithmic}[1]
			\Procedure{Placement}{$\mathbf{P},\mathcal{W}$}       
			\State Split each file $W_n$ in $\mathcal{W}$ into $F$ packets: $W_n =\{W_{n,j}: j \in [F]\}$
			\For{\texttt{$k \in [K]$}}
			\State  $\mathcal{Z}_k$ $\leftarrow$ $\{W_{n,j}: p_{j,k}=\star, \forall n \in [N]\}$
			\EndFor
			\EndProcedure
			
			\Procedure{Delivery}{$\mathbf{P},\mathcal{W},\phi,\vec{d}$} 
			\For{\texttt{$s \in [S]$}}
			\State User $U_{\phi(s)}$ sends $\underset{p_{j,k}=s, j\in [F],k\in[K]}{\bigoplus}W_{d_k,j}$
			\EndFor    
			\EndProcedure
		\end{algorithmic}
	\end{algorithm}
	Example \ref{eg:dpda} illustrates Algorithm \ref{d2d_alg} and Theorem \ref{thm:d2d}.
	\begin{example}\label{eg:dpda}
		It is easy to see that the following array $\mathbf{P}$ is a $(4,4,2,4)$ DPDA. The mapping $\phi(s)$ here is the identity function.  
		\begin{equation*}	
			\mathbf{P}	=	\begin{array}{cc|*{4}{c}}
				& & 1 & 2 & 3	& 4  \\
				\hline
				&	1 & \star & 3 &\star & 1 \\	
				&	2 & 3 & \star & \star & 2 \\
				&	3 & \star & 4 & 1 & \star \\
				&	4 & 4 & \star & 2 & \star  \\
			\end{array}
		\end{equation*}
		Based on this DPDA $\mathbf{P}$, one can obtain a $4$-division $(4,2,4)$ D2D coded caching scheme using Algorithm \ref{d2d_alg} as follows. \\
		\noindent $\bullet$ \textbf{Placement phase:} From line $2$ of Algorithm \ref{d2d_alg}, each file $W_n, \forall n \in [4]$ is split into $4$ packets, i.e., $W_n =\{W_{n,1},W_{n,2},W_{n,3},W_{n,4}\}$.  By line $3-5$ of Algorithm \ref{d2d_alg}, each user caches the packets as follows:
		\begin{equation*}
			\begin{split}
				& \mathcal{Z}_{1} = \{W_{n,1},W_{n,3} \}, \mathcal{Z}_{2} = \{W_{n,2},W_{n,4} \}, \\ & \mathcal{Z}_{3} = \{W_{n,1},W_{n,2} \},  \mathcal{Z}_{4} = \{W_{n,3},W_{n,4} \}, \forall n \in [4]. \\ 
			\end{split}
		\end{equation*}
		\noindent $\bullet$ \textbf{Delivery phase}: Let the demand vector $\vec{d}=\{4,2,1,3\}$. By line $8-10$ of Algorithm \ref{d2d_alg}, users $1$ to $4$   transmits $W_{3,1} \oplus W_{1,3}, W_{3,2} \oplus W_{1,4}, W_{4,2} \oplus W_{2,1}$ and $W_{4,4} \oplus W_{2,3}$ respectively. Each user can thus recover the requested file. For example, consider user $1$ whose demanded file is $W_4$. It has in its cache the packets $W_{4,1}$ and $W_{4,3}$ of the requested file $W_4$. From the transmission $W_{4,2} \oplus W_{2,1}$ by user $3$, user $1$ can recover $W_{4,2}$ since it has $W_{2,1}$ in its cache. From the transmission $W_{4,4} \oplus W_{2,3}$ by user $4$, user $1$ can recover $W_{4,4}$ since it has $W_{2,3}$ in its cache.  Thus user $1$ gets all the $4$ packets of the requested file $W_4$.
		
		There are $4$ transmissions in this example, one by each user, with each transmission being the size of a packet. Therefore $R=\frac{4}{4}=1$ files as mentioned in Theorem \ref{thm:d2d}.
	\end{example} 
	
	 \section{Multiaccess D2D coded caching}\label{prelim_madcc}
	In this section, we describe the MADCC network model and define the arrays used to characterize the MADCC schemes.
	 
\begin{figure}[H]
		\centering
		\captionsetup{justification=centering}
		\includegraphics[width=0.42\textwidth]{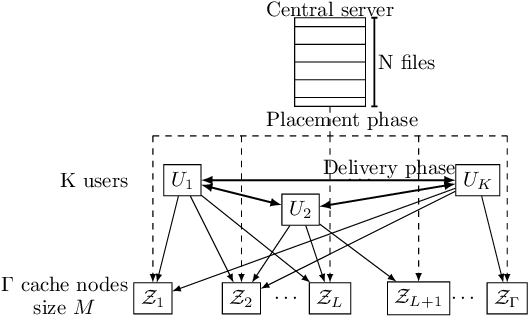}
		\caption{Multiaccess D2D coded caching network model.}
		\label{fig:setting_d2dmacc}
	\end{figure}

	An $(L,K,\Gamma,M,N)$ multiaccess D2D coded caching network with access topology $\mathfrak{B}$, depicted in Fig.\ref{fig:setting_d2dmacc}, consists of a central server with a library of $N$ files $\{W_n: n \in [N]\}$ each of size $B$ bits, $K \le N$ cache-less users $\{U_k: k \in [K]\}$ and $\Gamma$ cache nodes $\{\mathcal{Z}_\gamma: \gamma \in [\Gamma]\}$, each having a cache of size $M$ files, where $M\le \frac{N}{L}$. The set of cache nodes accessible to the user $U_k$ is denoted by $\mathcal{B}_k$, and each user can access a distinct set of $L$ cache nodes, i.e., $|\mathcal{B}_k|=L$, without any access cost. The access topology $\mathfrak{B}=\{\mathcal{B}_k\ |\ k\in[K]\}$. The condition $M\geq \frac{N}{LK}$ is assumed to ensure that any demands can be met using the cache contents.
	
	An MADCC scheme also operates in two phases. \\
		\noindent $\bullet$ \textbf{Placement phase}: During this phase, each file $W_{n}$ is split into $F$ non-overlapping packets. Each cache node $\mathcal{Z}_\gamma$, where $\gamma \in [\Gamma]\}$, caches $MB$ bits with \textit{uncoded data placement}. The later demands of the users are unknown during this phase.  \\
		\noindent $\bullet$ \textbf{Delivery phase}:  During this phase, each user requests a file from $\{W_n: n \in [N]\}$. For a given demand vector $\vec{d}$, using the cache contents it can access, each user broadcasts a coded message $X_{k,\vec{d}}$ consisting of $S_{k,\vec{d}}$ packets to all other users over an error-free shared medium. The delivery is assumed to be \textit{one-shot}. Through these transmissions and the accessible cache contents, all the users retrieve their requested file. 
		
Our objective is to design MADCC schemes with both the load per user and the subpacketization level as small as possible. Similar to designing DPDAs to obtain low-subpacketization D2D coded caching schemes, designing combinatorial structures that characterize the placement and delivery strategies of an MADCC scheme is meaningful in this direction. To this end, we define two arrays, the \textit{placement array} and the \textit{delivery array}. The placement array is defined to characterize data placement in the caches, and the delivery array is defined to characterize the user accessible cache contents and delivery strategy of the proposed MADCC schemes. These arrays are defined as follows.

\begin{defn}
		\noindent $\bullet$ (Placement array). An $F\times \Gamma$ placement array $\mathbf{Z}$ consists of  $\star$ and null, where $F$ and $\Gamma$ represent the subpacketization level and the number of cache-nodes, respectively. For any integers $f\in [F]$ and $\gamma\in [\Gamma]$, the entry $\mathbf{Z}(f,\gamma)$ is $\star$ if and only if the $\gamma^{\text{th}}$ cache node caches the $f^{\text{th}}$ packet of each $W_n$ where $n\in [N]$. Let the number of $\star$s in a column be $Z'$, then the cache memory ratio $\frac{M}{N}=\frac{Z'}{F}$. \\
		\noindent $\bullet$ (Delivery array). An $F\times K$ delivery array $\mathbf{Q}$ consists of elements from $\star \cup[S]$, where $F$ and $K$ represent the subpacketization level and the number of users, respectively. For any integers $f\in [F]$ and $k \in [K]$, the entry $\mathbf{Q}(f,k)$ is a $\star$ if and only if the $k^{\text{th}}$ user can retrieve the $f^{\text{th}}$ packet  of each $W_n$ from its connected cache nodes. Each integer $s \in [S]$ represents one multicast message by a user $\phi(s)$, and $S$ represents the total number of multicast messages sent in the delivery phase. To ensure that each user's demand is satisfied, the delivery array $\mathbf{Q}$ should be a DPDA. Therefore, the transmission load of the MADCC scheme is $R=\frac{S}{F}$. 
\end{defn}
\section{Combinatorial Designs }\label{designs}
Some basic useful definitions from combinatorial designs are described in this section. For further reading on this, readers can refer to \cite{Stin} and \cite{Col}.
\begin{defn}[Design $(\mathcal{X}, \mathcal{A})$]\cite{Stin}
	A design is a pair  $(\mathcal{X}, \mathcal{A})$ such that the following properties are satisfied: \\
	\textit{1}. $\mathcal{X}$ is a set of elements called points, and \hspace{0.1cm}
	\textit{2}. $\mathcal{A}$ is a collection (i.e., multiset) of nonempty subsets of $\mathcal{X}$ called blocks.
\end{defn}
\begin{defn}[$t-(v, k, \lambda)$ design]\cite{Stin}\label{def_tdes}
	Let $v,k,\lambda$ and $t$ be positive integers such that $v>k\geq t$. A $t-(v, k, \lambda)$ design is a design $(\mathcal{X}, \mathcal{A})$ such that the following properties are satisfied: \\
	\textit{1}. $|\mathcal{X}|=v$, \hspace{0.5cm}
	\textit{2}. each block contains exactly $k$ points, and \\
	\textit{3}. every set of $t$ distinct points is contained in exactly $\lambda$ blocks.
\end{defn}
A $t$-design with $t=2$ is called a balanced incomplete block design (BIBD). Large classes of BIBDs and general $t$-designs are given in \cite{Stin} and \cite{Col}.
For convenience, we write blocks of a $t-(v, k, \lambda)$ design in the form $abc$ rather than $\{a,b,c\}$.
\begin{example}\label{eg:tdes}
$\mathcal{X}=\{1,2,3,4,5,6,7\}$, $ \mathcal{A}=\{124,235,346, 457,156,267,137\}$ is a $2-(7,3,1)$ design.
\end{example}
The following theorems are useful in the construction of MADCC schemes with $t$-design topology in Section \ref{madcc_schemes_tdes}.
\begin{thm}\label{thm:Stin}\cite{Stin}
	 Let $(\mathcal{X}, \mathcal{A})$ is a $t-(v, k, \lambda)$ design. Suppose that $\mathcal{Y} \subseteq \mathcal{X}$, where $|\mathcal{Y}|=s\leq t.$ Then there are exactly $\lambda_{s}=\frac{\lambda \binom{v-s}{t-s}}{\binom{k-s}{t-s}}$ blocks in $\mathcal{A}$ that contains all the points in $\mathcal{Y}$. Thus the number of blocks in $\mathcal{A}$, $b=\lambda_{0}=\frac{\lambda \binom{v}{t}}{\binom{k}{t}}$.
\end{thm}
\begin{thm}\label{thm:tdes2}\cite{Stin}
	Let $(\mathcal{X}, \mathcal{A})$ be a $t-(v, k, \lambda)$ design. Suppose that $\mathcal{Y}, \mathcal{Z} \subseteq \mathcal{X}$, where $\mathcal{Y} \cap \mathcal{Z}=\emptyset, |\mathcal{Y}|=i, |\mathcal{Z}|=j$, and $i+j \le t$. Then there are exactly $\lambda_{i}^{j}=\frac{\lambda \binom{v-i-j}{k-i}}{\binom{v-t}{k-t}}$ blocks in $\mathcal{A}$ that contains all the points in $\mathcal{Y}$ and none of the points in $\mathcal{Z}$. 
\end{thm}

Next we review two other combinatorial structures, orthogonal arrays and group divisible designs, which we will use in our MADCC schemes with $t$-GDD topology in Section \ref{madcc_schemes_tgdd}.
\begin{defn}[Orthogonal Array]\cite{Stin}
	Let $t,v,r$ and $\lambda$ be positive integers such that $r\geq t \geq 2$. A $t-(v, r, \lambda)$ orthogonal array (denoted as $t-(v, r, \lambda)$-OA) is a pair $(\mathcal{X}, \mathbf{D})$ such that the following properties are satisfied: \\
	\textit{1}. $\mathcal{X}$ is a set of $v$ elements called points, \\
	\textit{2}. $\mathbf{D}$ is a $\lambda v^t \times r$ array whose entries are chosen from $\mathcal{X}$, \\
	\textit{3}. within any $t$ columns of $\mathbf{D}$, every $t$-tuple of points is contained in exactly $\lambda$ rows.
\end{defn}
\begin{example}\label{ex:oa}
${\small\mathcal{X}=\{0,1\}, {\setlength{\arraycolsep}{2pt}
		\mathbf{D}=\hspace{-0.1cm}\left(\begin{array}{ccc}
				1 & 1 & 0 \\
				0 & 0 & 0 \\
				1 & 0 & 1 \\
				0 & 1 & 1
			\end{array}\right)}
	} \text{ is a }2-(2,3,1)-OA.$
\end{example}

In \cite{XMCL}, it was shown that for any positive integers $m$ and $q$, where $m \ge 2$ and $q \ge 2$, there always exists an $(m-1)-(q,m,1)$ OA. This result is stated in the following lemma.
\begin{lem}\label{triv_OA}\cite{XMCL}
	For any $m \geq 2 $ and $q \geq 2$, there exist a $(m-1)-(q,m,1)$ OA $(\mathcal{Y}, \mathbf{D})$, in which $\mathcal{Y}=[q]$ and the rows of $\mathbf{D}$ are the vectors given by $\left\{ \left(d_1,d_2,d_3,...,d_{m-1}, d_m \right) :  \left(d_1,d_2,..., d_{m-1} \right) \in [q]^{m-1}, \right. \\ \left. \sum_{i=1}^{m} d_i \pmod{q} \text{ is a constant } \right\}$.
\end{lem}
Next, we review the definition of \textit{maximum distance separable codes (MDS codes)}. An $(n,M,d,q)$-code $\mathcal{C}$ is a non-empty subset $\mathcal{C} \subseteq \mathbb{F}_q^n$, with $|\mathcal{C}|=M$ and minimum distance $d$. An $(n,M,d,q)$-code $\mathcal{C}$ in which $ M = q^{n-d+1}$ is called an \textit{MDS code} \cite{Stin}. An equivalence between orthogonal arrays with $\lambda=1$ and MDS codes is stated in the following theorem.
\begin{thm}\cite{Stin}\label{thm:oa_mds}
	An $(n,M,d,q)$ MDS code $\mathcal{C}$, is equivalent to a $()n-d+1)-(q,n,1)$ OA $(\mathcal{X}, \mathbf{D})$, where the codewords of $\mathcal{C}$ are the $v^t$ rows of $\mathbf{D}$.
\end{thm}
The following lemma follows from Theorem \ref{thm:oa_mds}.
\begin{lem}\label{lem:oa_mds}
For a $t-(v, r, 1)$-OA $(\mathcal{X}, \mathbf{D})$, $d_{min}(\mathbf{D})\triangleq \min\limits_{i,j \in [v^t], i \ne j  } d\left(\vec{D}_i,\vec{D}_j\right)=r-t+1$, where $\vec{D}_i$ denotes the $i^{th}$ row of $\mathbf{D}$. 
\end{lem}

\begin{defn}[$t$-GDD]\cite{HM}
	Let $v$ be a non-negative integer, $\lambda$ and $t$ be positive integers and $K$ be a set of positive integers.  A $t$-group divisible design ($t$-GDD) of order $v$,index $\lambda$ and block sizes from $K$ is a triple $(\mathcal{X}, \mathcal{G}, \mathcal{A})$ where \\
	\textit{1}. $\mathcal{X}$ is a set of $v$ elements called points, \hspace{0.2cm}
	\textit{2}. $\mathcal{G}=\{G_1,G_2,...\}$ is a non-empty subsets of $\mathcal{X}$ which partition $\mathcal{X}$ called groups, \\
	\textit{3}. $\mathcal{A}$ is a family of subsets of $\mathcal{X}$ each of cardinality from $K$ called blocks such that each block intersects any given group in at most one point, and \hspace{0.2cm}
	\textit{4}. each $t$-set of points from $t$ distinct groups is in exactly $\lambda$ blocks.
\end{defn}
The above definition of $t$-GDD is a generalization of the concept of group divisible design discussed in \cite{Stin} for $t \geq 2$, with the latter corresponding to the case where $t=2$. For more details on $t$-GDDs, readers can refer to \cite{Stin}, \cite{Col} and \cite{HM}. 
We denote a $t$-GDD with block size $k$ and $m$ groups of uniform group size $q$ (therefore $v=mq$) as $t-(m,q,k,\lambda)$ GDD. We consider  $t-(m,q,k,\lambda)$ GDDs as access topology for our MADCC schemes in Section \ref{madcc_schemes_tgdd}. In a $t-(m,q,k,\lambda)$ GDD, the number of blocks in $\mathcal{A}$ is $\frac{\lambda\binom{m}{t}q^t}{\binom{k}{t}}$ \cite{SHMWT}.
\begin{example}\label{eg:GDD}
$\mathcal{X}=\{1,2,3,4,5,6\}$, $\mathcal{G}=\{G_1=\{1,2\}, G_2=\{3,4\},G_3=\{5,6\}\}$ and $\mathcal{A}=\{\{1,3,5\}, \{2,3,6\}, \{1,4,6\}, \\ \{2,4,5\}\}$ is a $2-(3,2,3,1)$ GDD.
\end{example}
\begin{thm}\label{thm:tgdd}\cite{SHMWT}
	Suppose that $(\mathcal{X}, \mathcal{G}, \mathcal{A})$ is a $t-(m,q,k,\lambda)$ GDD. Then, for any $i \in [t]$, a $t-(m,q,k,\lambda)$ GDD is also a $i-(m,q,k,\lambda_{i})$ GDD with $\lambda_{i}=\frac{\lambda q^{t-i} \binom{m-i}{t-i}}{\binom{k-i}{t-i}}$.
\end{thm}

The point representation of a $t-(m,q,k,\lambda)$ GDD is obtained by denoting the $v^{th}$ point of the $u^{th}$ group by a vector $(u,v)$, for any $u \in [m]$ and $v \in [q]$ \cite{MACC_des}.

One can construct a $t$-GDD for any integers $t$, $m$ and $q$ with $1<t<m$, and $q \geq 2$, as stated in the following lemma.
\begin{lem}\label{triv_GDD}
	For any positive integers $t$ and $m$ with $1<t<m$, and $q \geq 2$, there always exists a $t-(m,q,t,1)$ GDD.
\end{lem}
\begin{IEEEproof}
	Consider any positive integers $t$ and $m$ with $1<t<m$, and $q \geq 2$. Take  $\mathcal{X}=\{(u,v) : u \in [m], v \in [q]\}$, $\mathcal{G}=\{G_1,G_2,...,G_m\}$ where $G_u=\{(u,1), (u,2),...,(u,q)\}$, $\forall u \in [m]$ and $\mathcal{A}$ is the collection of all $t$-set of points selected from $t$ distinct groups among the $m$ groups. That is, $\mathcal{A}=\{A_1,A_2,..,A_{\binom{m}{t}q^t}\} :A_i=\{(u_{i_1},v_{i_1}), (u_{i_2},v_{i_2}),\\...,(u_{i_t},v_{i_t})\}$, $u_{i_1} \ne u_{i_2} \ne ...\ne u_{i_t}, u_{i_j} \in [m], v_{i_j}\in [q], \forall j \in [t], \forall i \in [\binom{m}{t}q^t]$. It is easy to see that $(\mathcal{X}, \mathcal{G}, \mathcal{A})$ is a $t-(m,q,t,1)$ GDD.
\end{IEEEproof}
\section{MADCC schemes with $t$-design topology}\label{madcc_schemes_tdes}
In this section, we propose a novel MADCC scheme in which the access topology is described by a $t-(v,k,\lambda)$ design, and obtain the derived CWEC scheme with the same topology. We then compare the proposed scheme with the derived CWEC scheme and the WCCWC scheme. The idea is to represent the caches in the MADCC network by the points of a given $t$-design, while the users are represented by the blocks of the $t$-design. A user denoted by a block can access all the caches represented by the points in that block. Property \textit{2} in Definition \ref{def_tdes} ensures that each user has access to the same number of caches. Property \textit{3} in Definition \ref{def_tdes} and the properties mentioned in Theorem \ref{thm:Stin} and Theorem \ref{thm:tdes2} make $t$-designs particularly suitable to construct the placement array and the delivery array of the proposed MADCC scheme. Before discussing the scheme, we first define the term \textit{$t$-design topology}.

\begin{defn}[$t$-design topology]\cite{MACC_des}
	Let $(\mathcal{X}, \mathcal{A})$ be a $t-(v,k,\lambda)$ design. A $t$-design topology is defined as an access topology $\mathfrak{B}$ in which \\
	\textit{1}. the users are represented by the blocks $A \in \mathcal{A}$, \\
	\textit{2}. the cache nodes are represented by the points $x \in \mathcal{X}$, \\
	\textit{3}. and the user $\{U_A : A \in \mathcal{A}\}$ can access all the caches $x$ such that $x \in A$, i.e., $\mathfrak{B}=\mathcal{A}$.
\end{defn}

\subsection{MADCC schemes with $t$-design topology}
In this subsection, we first propose a novel $(k,\frac{\lambda \binom{v}{t}}{\binom{k}{t}},v,M,N)$ MADCC scheme with $t$-design topology given by Theorem \ref{thm:MADCC_tdes}. We then obtain the derived CWEC scheme with $t$-design topology. The MADCC schemes are characterized by the placement array and the delivery array, as defined in Section \ref{prelim_madcc}.
\begin{thm}\label{thm:MADCC_tdes}
	Given a $t-(v,k,\lambda)$ design $(\mathcal{X}, \mathcal{A})$, there exists a $\binom{v}{i}\binom{k}{t-i}$-division $(k,\frac{\lambda \binom{v}{t}}{\binom{k}{t}},v,M,N)$ multiaccess D2D coded caching system with access topology $\mathfrak{B}=\mathcal{A}$, for which the following memory-load points are achievable,
	\begin{equation*}
		\left(\frac{M}{N},R\right)=\left(\frac{i}{v}, \frac{\lambda\binom{v-i}{k}}{\binom{v-t}{k-t}\binom{t}{i}}\right),  \text{ for each } i \in [t-1].
	\end{equation*}
\end{thm}
\begin{IEEEproof}
	Given a $t-(v,k,\lambda)$ design $(\mathcal{X}, \mathcal{A})$, the cache nodes are represented by the points in $\mathcal{X}$, therefore, $\Gamma=v$. The users are represented by the blocks $A \in \mathcal{A}$, therefore, $K=\frac{\lambda\binom{v}{t}}{\binom{k}{t}}$. The access topology $\mathfrak{B}=\mathcal{A}$, i.e., the user $\{U_A : A \in \mathcal{A}\}$ can access all the caches $\{x \in \mathcal{X} : x \in A\}$. Therefore, $L=|A|=k$. Now we describe our MADCC scheme by constructing the placement array and the delivery array. \\ \\
	\textit{I. Placement array}
	
	  The central server first splits each file $\{W_n : n \in [N]\}$ into $\binom{v}{i}$, where $i \in [t-1]$, non-overlapping packets, i.e., $\{W_n=W_{n,\mathcal{D}} : \mathcal{D} \in \binom{[v]}{i}, n \in [N]\}$. We further divide each packet $W_{n,\mathcal{D}}$ into  $\binom{k}{t-i}$ sub-packets, i.e., $\{W_{n,\mathcal{D}}=W_{n,\mathcal{D}}^{\mathcal{T}} : \mathcal{T} \in \binom{[k]}{t-i}, \mathcal{D} \in \binom{[v]}{i}, n \in [N]\}$. Therefore, $F=\binom{v}{i}\binom{k}{t-i}$. Now we construct a $\binom{v}{i}\binom{k}{t-i} \times v$ placement array $\mathbf{Z}=\left(Z_{(\mathcal{D},\mathcal{T}),x}\right)_{\mathcal{D} \in \binom{[v]}{i}, \mathcal{T} \in \binom{[k]}{t-i}, x \in \mathcal{X} }$, whose rows are indexed by $\left(\mathcal{D},\mathcal{T}\right)$ and columns are indexed by $x$, as follows.
	  \begin{equation}\label{eq:pa_tdes}
	  	Z_{(\mathcal{D},\mathcal{T}),x}= \begin{cases}  \star & \text {if } x \in \mathcal{D}   \\ null & \text {otherwise }.\end{cases}
	  \end{equation}
	  From (\ref{eq:pa_tdes}), it can be seen that the sub-packets cached in each cache node $x \in \mathcal{X}$ is $\mathcal{Z}_x=\{W_{n,\mathcal{D}}^{\mathcal{T}} : x \in \mathcal{D}, \mathcal{D} \in \binom{[v]}{i}, \mathcal{T} \in \binom{[k]}{t-i}, n \in [N]\}\}$. Therefore, the number of sub-packets of each file cached in each cache node, which is the same as the number of $\star$s in each column of $\mathbf{Z}$, is $Z'=\binom{v-1}{i-1}\binom{k}{t-i}$. Therefore, the cache memory ratio $\frac{M}{N}=\frac{Z'}{F}=\frac{\binom{v-1}{i-1}\binom{k}{t-i}}{\binom{v}{i}\binom{k}{t-i}}=\frac{i}{v}$. \\ \\
	  \textit{II. Delivery array}
	  
	  Each user $\{U_A : A \in \mathcal{A}\}$ can access the contents in the cache nodes $\{x : x \in A\}$. Therefore, the sub-packets that the user $U_A$ can retrieve, denoted by $\mathcal{Z}_A$, is $\{W_{n,\mathcal{D}}^{\mathcal{T}} : A \cap \mathcal{D} \ne \emptyset, \mathcal{D} \in \binom{[v]}{i}, \mathcal{T} \in \binom{[k]}{t-i}, n \in [N]\}\}$.   Since the number of users is $\frac{\lambda \binom{v}{t}}{\binom{k}{t}}$, we construct a  $\binom{v}{i}\binom{k}{t-i} \times \frac{\lambda \binom{v}{t}}{\binom{k}{t}}$ delivery array $\mathbf{Q}=\left(Q_{(\mathcal{D},\mathcal{T}),A} \right)_{\mathcal{D} \in \binom{[v]}{i}, \mathcal{T} \in \binom{[k]}{t-i}, A \in \mathcal{A} }$, as follows.
	   \begin{equation}\label{eq:da_tdes}
	  	Q_{(\mathcal{D},\mathcal{T}),A}= \begin{cases}  \star & \text {if } A \cap \mathcal{D} \ne \emptyset   \\ \left(\mathcal{D} \cup A(\mathcal{T})\right)_{\alpha(\mathcal{D},A(\mathcal{T}))} & \text {otherwise, }\end{cases}
	  \end{equation}
	  where $A(\mathcal{T})$ represents the points in $A$ with index in $\mathcal{T}$ and $\alpha(\mathcal{D},A(\mathcal{T}))$ is the occurrence number of $\mathcal{D} \cup A(\mathcal{T})$ in $\mathbf{Q}$ from top to bottom and left to right for a given $\mathcal{D}$.
	  \begin{claim}
	  	\label{claim:dpda_tdes}
	  	The delivery array $\mathbf{Q}$ is in fact a $\left(\frac{\lambda \binom{v}{t}}{\binom{k}{t}}, \binom{v}{i}\binom{k}{t-i}, \left(\binom{v}{i}-\binom{v-k}{i}\right)\binom{k}{t-i}, \frac{\lambda\binom{v}{t}\binom{v-t}{k-t+i}}{\binom{v-t}{k-t}} \right)$ DPDA.
	  \end{claim} 
	  Proof of Claim \ref{claim:dpda_tdes} is given in \textit{Appendix \ref{appendix:dpda_tdes}}. Since $\mathbf{Q}$ is a DPDA, each user's demand can be satisfied with a transmission load $R=\frac{S}{F}=\frac{\lambda\binom{v}{t}\binom{v-t}{k-t+i}}{\binom{v-t}{k-t}\binom{v}{i}\binom{k}{t-i}}= \frac{\lambda\binom{v-i}{k}}{\binom{v-t}{k-t}\binom{t}{i}}$.
\end{IEEEproof}
The following example illustrates Theorem \ref{thm:MADCC_tdes}.
\begin{example}\label{ex:madcc_tdes}
From the $2-(7,3,1)$ design in Example \ref{eg:tdes} and for $i=1$, a $21$-division $(3,7,7,1,7)$ MADCC scheme with access topology $\mathfrak{B}=\mathcal{A}$ is obtained as follows. The $7$ cache nodes are represented by the $7$ points in $\mathcal{X}$, and the $7$ users are represented by the $7$ blocks in $\mathcal{A}$, and each user can access $3$ cache nodes that are represented by the $3$ points in the corresponding block. Each file is divided into $7 \times 3 =21$ sub-packets, i.e., $\{W_{n}=W_{n,\mathcal{D}}^{\mathcal{T}} : \mathcal{D} \in [7], \mathcal{T} \in [3], \forall n \in [7]\}$. Then by (\ref{eq:pa_tdes}), we obtain the $21 \times 7$ placement array shown in Fig.\ref{fig:pa_tdes}. 
\begin{figure}[!htbp]
	\centering
	\begin{subfigure}{0.38\textwidth} 
		\centering
		\includegraphics[width=0.8\linewidth]{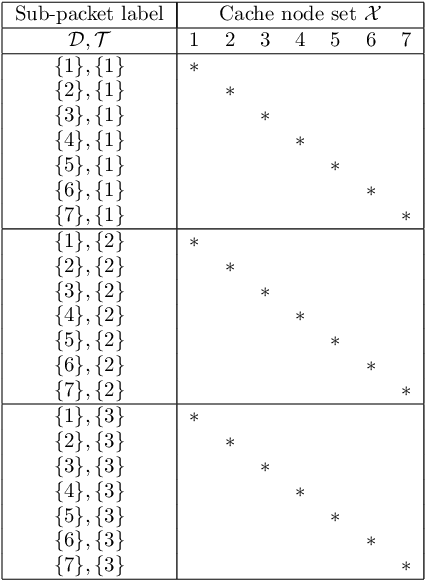}
		\caption{Placement array}
		\label{fig:pa_tdes}
	\end{subfigure}
	\hfill
	\begin{subfigure}{0.49\textwidth} 
		\centering
		\includegraphics[width=0.9\linewidth]{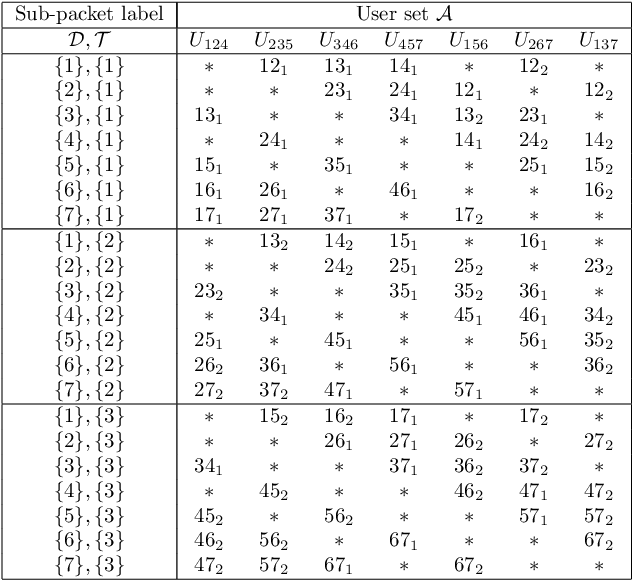} 
		\caption{Delivery array}
		\label{fig:da_tdes}
	\end{subfigure}
	\caption{Placement array and delivery array in Example \ref{ex:madcc_tdes}.}
\end{figure}

By (\ref{eq:da_tdes}), we obtain the $21 \times 7$ delivery array shown in Fig.\ref{fig:da_tdes}. 
To make clear how the subscript $\alpha(\mathcal{D},A(\mathcal{T}))$ is written, consider the case $\mathcal{D}=\{1\}$. From Theorem \ref{thm:tdes2}, it is known that the number of blocks in $\mathcal{A}$ that contain any subset of size $t-1=1$ but do not contain the point $\mathcal{D}=\{1\}$ is $\lambda_1^1= 2$. Therefore, for $\mathcal{D}=\{1\}$ and , $\forall \mathcal{T}=\{1\}, \{2\}$ and $\{3\}$, a specific $\mathcal{D}\cup A(\mathcal{T})=\{1\} \cup A(\mathcal{T})$ appears twice all together in the rows $\{1\},\{\mathcal{T} \}$. Therefore, $\alpha(\mathcal{D},A(\mathcal{T}))$ can take values $1$ and $2$ depending on the occurrence of $\{1\} \cup A(\mathcal{T})$ from top to bottom and left to right in $\mathbf{Q}$.

It can be verified that the delivery array in Fig.\ref{fig:da_tdes} is a $(7,21,9,42)$ DPDA and therefore, using Algorithm \ref{d2d_alg}, one can obtain the scheme with load $R=\frac{S}{F}=2$ and subpacketization level $F=21$. The mapping from non-$\star$ entries to the users is $\phi(ab_\alpha)= U_A \text{ such that } \{ab\} \subset A $. Since every $t=2$ sized subset appears in exactly one block, there is a unique user corresponding to each non-$\star$ entry, and each user broadcasts $\binom{3}{2}\times2=6$ messages. 
\end{example}

By slightly modifying the MADCC scheme in Theorem \ref{thm:MADCC_tdes}, we can obtain the following MADCC scheme in Corollary \ref{cor:MADCC_tdes}.
\begin{cor}\label{cor:MADCC_tdes}
	Given a $t-(v,k,\lambda)$ design $(\mathcal{X}, \mathcal{A})$, it is possible to get a $\binom{v}{i}\binom{k}{t-1}$-division $(k,\frac{\lambda \binom{v}{t}}{\binom{k}{t}},v,M,N)$ multiaccess D2D coded caching system with access topology $\mathfrak{B}=\mathcal{A}$, for which the following memory-load points are achievable,
	${\Large
		\left(\frac{M}{N},R\right)=\left(\frac{i}{v}, \frac{\lambda\binom{v-i}{k}}{\binom{v-t}{k-t}\binom{i+t-1}{i}}\right) }
	$, for each $i \in [\min(t,k-t+1)]$, provided for any $\mathcal{Y} \subseteq A \in \mathcal{A}$ with $t < |\mathcal{Y}| \le k$, the number of blocks in $\mathcal{A}$ that contain $\mathcal{Y}$, $\lambda(\mathcal{Y})$, is same for all $\mathcal{Y}$. 
\end{cor}
Proof of Corollary \ref{cor:MADCC_tdes} is given in Appendix \ref{appendix:dpda_tdes_cor}. 
 	\begin{figure*}[!htbp]
 	\centering
 	\begin{subfigure}{0.3\textwidth}
 		\centering
 		\includegraphics[width=0.85\linewidth]{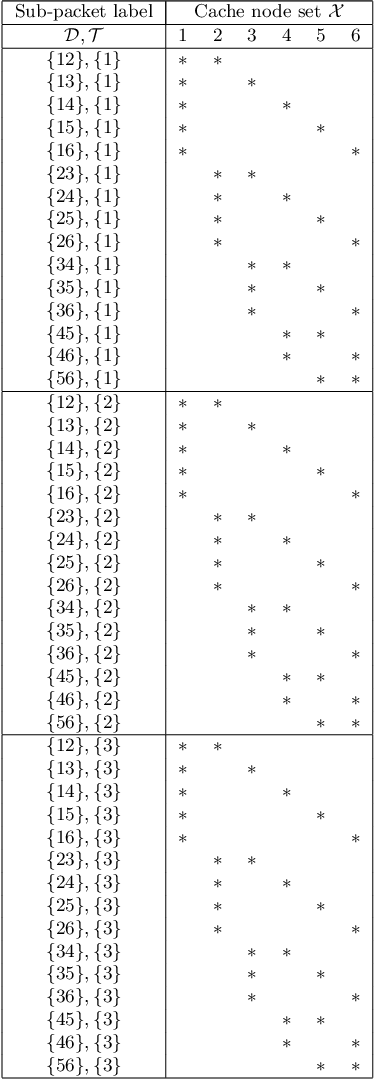}
 		\caption{Placement array}
 		\label{fig:pa_tdes_cor1}
 	\end{subfigure}%
 	\hspace{0.5cm} 
 	\begin{subfigure}[b]{0.63\textwidth}
 		\centering
 		\includegraphics[width=0.85\linewidth]{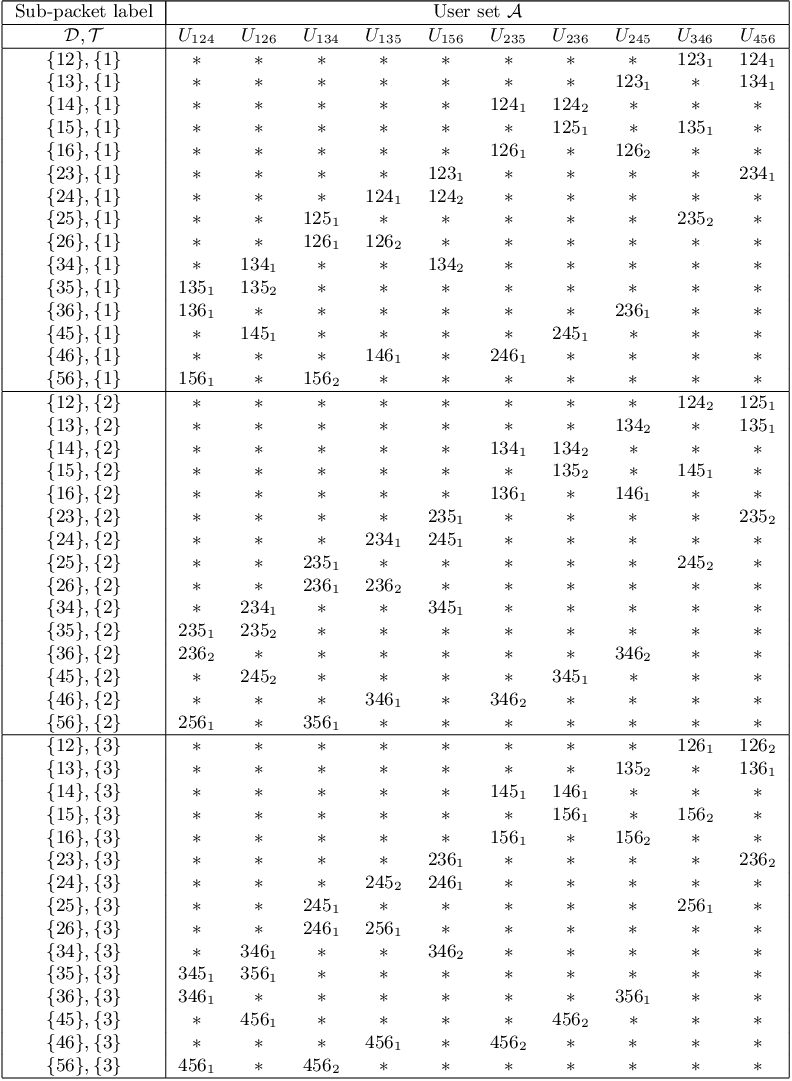}
 		\caption{Delivery array}
 		\label{fig:da_tdes_cor1}
 	\end{subfigure}
 	\caption{Placement array and delivery array in Example \ref{ex:madcc_tdes_cor1}.}
 \end{figure*}
The following example illustrates Corollary \ref{cor:MADCC_tdes}.
\begin{example}\label{ex:madcc_tdes_cor1}
	$\mathcal{X}=\{1,2,3,4,5,6\}$, $ \mathcal{A}=\{124,126,134,135,156,235,236,245,346,456\}$ is a $2-(6,3,2)$ design. From this design and for $i=2$, a $45$-division $(3,10,6,2,6)$ MADCC scheme with access topology $\mathfrak{B}=\mathcal{A}$ is obtained as follows. The $6$ cache nodes are represented by the $6$ points in $\mathcal{X}$, and the $10$ users are represented by the $10$ blocks in $\mathcal{A}$, and each user can access $3$ cache nodes that are represented by the $3$ points in the corresponding block. Each file is divided into $\binom{6}{2} \times \binom{3}{1} =45$ sub-packets, i.e., $\{W_{n}=W_{n,\mathcal{D}}^{\mathcal{T}} : \mathcal{D} \in \binom{[6]}{2}, \mathcal{T} \in [3], \forall n \in [6]\}$. Then by (\ref{eq:pa_tdes}), we obtain the $45 \times 6$ placement array shown in Fig.\ref{fig:pa_tdes_cor1}. By (\ref{eq:da_tdes}), we obtain the $45 \times 10$ delivery array shown in Fig.\ref{fig:da_tdes_cor1}. 	It can be verified that the delivery array in Fig.\ref{fig:da_tdes_cor1} is a $(10,45,36,30)$ DPDA. Here, $|\mathcal{D} \cup A(\mathcal{T})|=k=3$ and it can be seen that for any given $\mathcal{D}$, $\lambda_1^2=2$ if $\mathcal{T}$ is such that $\mathcal{D} \cup A(\mathcal{T}) = A_i \in \mathcal{A}$ and $\lambda_1^2=1$ if $\mathcal{T}$ is such that $\mathcal{D} \cup A(\mathcal{T}) \ne A_i \in \mathcal{A}$. Therefore, the number of distinct non-$\star$ entries $S= b \times 2 + \left(\binom{v}{3}-b\right) \times 1 = 10 \times 2 + (20-10) \times 1 =30$ as evident from the delivery array. Here, the mapping is $\{\phi\left(\left(\mathcal{D} \cup A(\mathcal{T})\right)_{\alpha(\mathcal{D},A(\mathcal{T}))}\right)= U_{A_i} : |\left(\mathcal{D} \cup A(\mathcal{T})\right) \cap A_i |\geq 2 \} $. Thus, here if  $\mathcal{D} \cup A(\mathcal{T}) = A_i \in \mathcal{A}$, there are $4$ users satisfying this condition, and any one of them can perform the transmission.  If $\mathcal{D} \cup A(\mathcal{T}) \ne A_i \in \mathcal{A}$, there are $6$ users satisfying this condition, and any one of them can perform the transmission. 
\end{example}

\begin{figure*}[!htbp]
	\centering
	\begin{subfigure}[t]{0.45\textwidth}
		\centering
		\includegraphics[width=\linewidth]{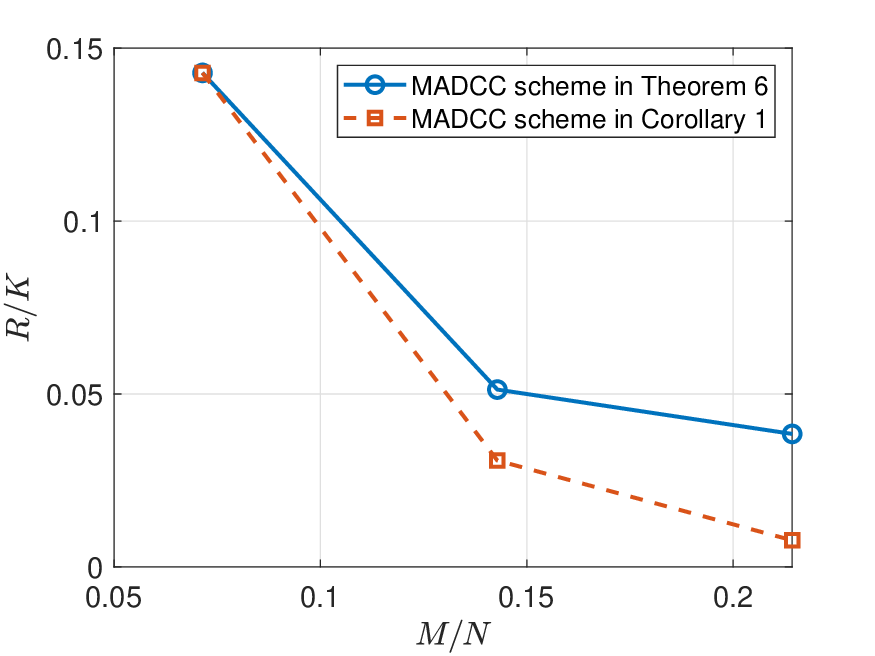}
		\subcaption{}
		\label{fig:madcc_tdes}
	\end{subfigure}
	\begin{subfigure}[t]{0.45\textwidth}
		\centering
		\includegraphics[width=\linewidth]{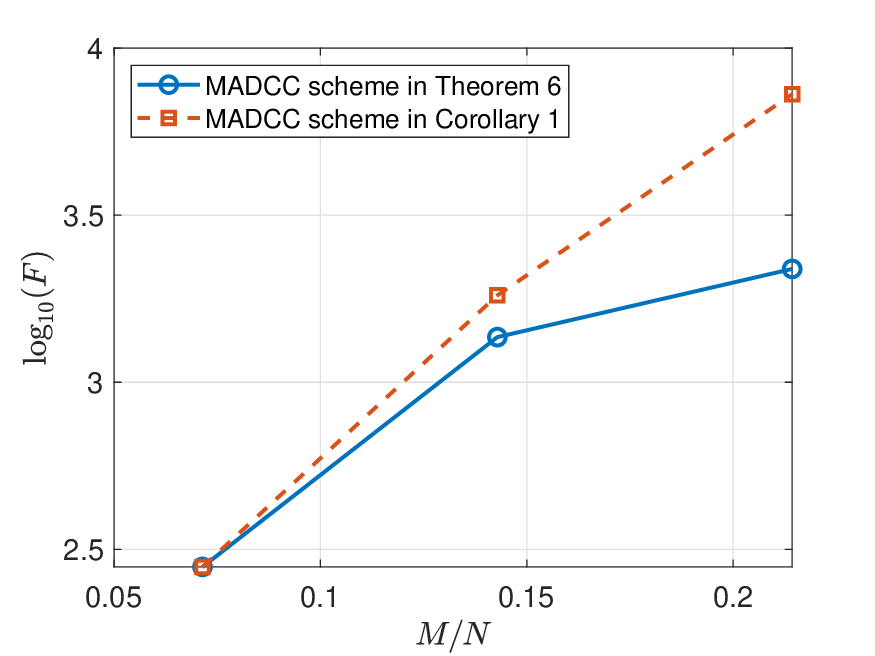}
		\subcaption{}
		\label{fig:madcc_tdes2}
	\end{subfigure}
	\caption{Load per user and subpacketization level as a function of memory of schemes in Theorem \ref{thm:MADCC_tdes} and Corollary \ref{cor:MADCC_tdes} for a $(6,1001,14,M,N)$ multiaccess D2D network.}
	\label{fig:madcc}
\end{figure*}

 Given a $t-(v,k,\lambda)$, Theorem \ref{thm:MADCC_tdes} and Corollary \ref{cor:MADCC_tdes} give MADCC schemes for a $(k,\frac{\lambda \binom{v}{t}}{\binom{k}{t}},v,M,N)$ multiaccess D2D network at $t-1$ and $\min(t,k-t+1)$ memory points, respectively. By memory sharing, the lower convex envelope of these obtained $(M, R)$ points is achievable. This is illustrated with an example by choosing a $4-(14,6,15)$ design (Table $4.37$, \cite{Col}). The load per user and the subpacketization level of these schemes at various memory points are plotted in Fig.\ref{fig:madcc_tdes} and Fig.\ref{fig:madcc_tdes2}, respectively.

\begin{table*}[!htbp]
	\begin{center}
		\resizebox{0.85\textwidth}{!}{
			\begin{tabular}{|c|c|c|c|c|c|c|c|}
				\hline
				\textbf{\makecell{Schemes and  parameters}} & \textbf{$K$} & \textbf{$L$} & \textbf{$\Gamma$} &\textbf{$\frac{M}{N}$} & \textbf{$R$} & \textbf{$\frac{R}{K}$} & \textbf{ $F$} \\\hline
				\makecell{Proposed scheme with $t$-design topology \\in Theorem \ref{thm:MADCC_tdes}, $t-(v,k,\lambda)$ design and $i \in [t-1]$}
				& $\frac{\lambda\binom{v}{t}}{\binom{k}{t}}$ & $k$ & $v$ & $\frac{i}{v}$ & $\frac{\lambda\binom{v-i}{k}}{\binom{v-t}{k-t}\binom{t}{i}}$ & $\frac{\binom{v-k}{i}}{\binom{v}{i}\binom{t}{i}}$ & $\binom{v}{i}\binom{k}{t-i}$ \\
				\hline
				\makecell{Derived CWEC scheme with $t$-design topology \\ in Lemma \ref{lem:derived_MADCC_tdes}, $t-(v,k,1)$ design}
				& $\frac{\binom{v}{t}}{\binom{k}{t}}$ & $k$ & $v$ & $\frac{1}{v}$ & $\frac{\binom{v-1}{k}}{\binom{v-t}{k-t}t}$ & $\frac{v-k}{vt}$ & $vt\binom{k}{t}$ \\
				\hline
				\makecell{WCCWC scheme in \cite{WCCWC}, integer $r=k>1$,\\  MAN PDA with $t_0=i$ and $K_0=\frac{v-i}{k-1} \in \mathbb{Z}^+ \backslash [i]$}
				& $v$ & $k$ & $v$ & $\frac{i}{v}$ & $\frac{v-ki}{i}$ & $\frac{v-ki}{vi}$ & $\binom{\frac{v-i}{k-1}}{i}vi$ \\
				\hline
		\end{tabular} }
	\end{center}
	\caption{Comparison between the proposed scheme in Theorem \ref{thm:MADCC_tdes}, the derived scheme in Lemma \ref{lem:derived_MADCC_tdes} and the WCCWC scheme.}
	\label{tab:MADCC_tdes}
\end{table*}

Next, we discuss the derived CWEC scheme with $t$-design topology. Given a $(g+1)-(K,F,Z,S)$ PDA, Wang \textit{et al.} in \cite{JMQX} proposed a method to construct a $(K,gF,gZ,(g+1)S)$ DPDA. Therefore, whenever the user-retrieve array (equivalent to the delivery array) designed for the MACC scheme with $t$-design topology in \cite{MACC_des} is a $(g+1)-(K,F,Z,S)$ regular PDA, it can be transformed into a $(K,gF,gZ,(g+1)S)$ DPDA to obtain the derived CWEC scheme with $t$-design topology, for the same system parameters. The placement array of this derived CWEC scheme will be the same as the node-placement array designed for the MACC scheme with $t$-design topology in \cite{MACC_des}, except that the number of sub-packets to which each packet is divided gets increased by a factor of $g$. For a $t-(v,k,1)$ design and $i=1$, for a cache memory $\frac{M}{N}=\frac{1}{v}$, the user-retrieve array in  \cite{MACC_des} is a $(t+1)-\left(\frac{ \binom{v}{t}}{\binom{k}{t}}, v\binom{k}{t}, k\binom{k}{t}, \binom{v}{t+1}-\frac{\binom{v}{t}}{\binom{k}{t}}\binom{k}{t+1} \right)$ PDA. Therefore the delivery array of the corresponding derived CWEC scheme with $t$-design topology is a $\left(\frac{ \binom{v}{t}}{\binom{k}{t}}, vt\binom{k}{t}, kt\binom{k}{t}, (t+1)\left( \binom{v}{t+1}-\frac{\binom{v}{t}}{\binom{k}{t}}\binom{k}{t+1} \right) \right)$ DPDA. In the resulting MADCC scheme, there are $v$ caches each having a memory $\frac{M}{N}=\frac{1}{v}$. The transmission load of the derived CWEC scheme with $t$-design topology is given by, $R=\frac{S}{F}=\frac{(t+1)\left( \binom{v}{t+1}-\frac{\binom{v}{t}}{\binom{k}{t}}\binom{k}{t+1} \right)}{vt\binom{k}{t}}=\frac{(t+1)\left(\binom{v}{t}\frac{v-k}{t+1}\right)}{vt\binom{k}{t}}= \frac{\binom{v-1}{k}}{\binom{v-t}{k-t}t}$. This scheme is stated as the following lemma.
 \begin{lem}\label{lem:derived_MADCC_tdes}
 	(Derived CWEC scheme with $t$-design topology) Given a $t-(v,k,1)$ design $(\mathcal{X}, \mathcal{A})$, there exist a $vt\binom{k}{t}$-division $(k,\frac{ \binom{v}{t}}{\binom{k}{t}},v,M,N)$ multiaccess D2D coded caching system with access topology $\mathfrak{B}=\mathcal{A}$, for which the  memory-load point 
 	$\left(\frac{M}{N},R\right)=\left(\frac{1}{v}, \frac{\binom{v-1}{k}}{\binom{v-t}{k-t}t}\right)
 	$ is achievable.
 \end{lem} 
\subsection{Performance analysis of the proposed MADCC scheme with $t$-design topology}
In this subsection, we carry out a comparison of the proposed MADCC scheme with $t$-design topology in Theorem \ref{thm:MADCC_tdes}, the derived CWEC scheme with $t$-design topology in Lemma \ref{lem:derived_MADCC_tdes} and the WCCWC scheme, using the metrics of load per user and subpacketization level while keeping the number of caches, cache size and access degree the same. These schemes are compared in Table \ref{tab:MADCC_tdes}, considering $v$ number of caches and an access degree of $k$.

 From Table \ref{tab:MADCC_tdes} it is clear that, by choosing $\lambda=1$ and $i=1$ in the proposed scheme in Theorem \ref{thm:MADCC_tdes}, the system parameters of the proposed scheme become identical to those of the derived CWEC scheme in Lemma \ref{lem:derived_MADCC_tdes}.  It is observed that our proposed scheme in Theorem \ref{thm:MADCC_tdes} outperforms the derived CWEC scheme in Lemma \ref{lem:derived_MADCC_tdes} in terms of subpacketization level, while achieving the same load per user. That is, $\frac{F_{proposed}}{F_{derived}}=\frac{v\binom{k}{t-1}}{vt\binom{k}{t}}=\frac{1}{k-t+1}$.
 	
  Next, we compare the proposed scheme in Theorem \ref{thm:MADCC_tdes} and the WCCWC scheme, considering the same number of caches, access degree, and cache size, as shown in Table \ref{tab:MADCC_tdes}. By choosing $i=1$ in the proposed scheme and the WCCWC scheme, from Table \ref{tab:MADCC_tdes} it is seen that, $\frac{{\text{Load per user}}_{proposed}}{{\text{Load per user}}_{WCCWC}}=\frac{1}{t} < 1$. From the examples in rows $1$ to $6$ of Table \ref{tab:MADCC_tdes2}, it is clear that, for $i=1$, the subpacketization level of the proposed scheme is equal in some cases and varies (lesser or higher) in others. For $i \ge 2$, we obtain $\frac{{\text{Load per user}}_{proposed}}{{\text{Load per user}}_{WCCWC}}=\frac{\binom{v-k}{i}vi}{\binom{t}{i}\binom{v}{i}(v-ki)}$. It is hard to comment on this ratio in general. Therefore, we consider specific examples as given in Table \ref{tab:MADCC_tdes2}. For all even integers $v \ge 6$, there exists a $3-(v,4,3)$ design (Theorem $9.14$ \cite{Stin}). By choosing $3-(v,4,3)$ design with $v=6k+2, \forall k \ge 2$ and $i=2$ in the proposed scheme, and $r=k \ge 2$, $t_0=2$ and $K_0=2k$ in the WCCWC scheme, schemes are obtained with parameters specified in rows numbered $7$ and $8$ in Table \ref{tab:MADCC_tdes2}, respectively. For this case, we have $\frac{{\text{Load per user}}_{proposed}}{{\text{Load per user}}_{WCCWC}}=\frac{2(3k-1)(2k-1)}{3(6k+1)(k-1)} < 1$, $\forall k$. Also, $\frac{F_{proposed}}{F_{WCCWC}}=\frac{6k+1}{k(2k-1)} <  1$, $\forall k \ge 4$. Next, we consider another class of $t$-designs. For all $q$ such that $q \equiv 1 (mod \text{ } 3)$, there exists a $3-(q+1,4,2)$ design (Table $4.37$ \cite{Col}). By choosing $3-(q+1,4,2)$ design with $q$ such that $q \equiv 1 (mod \text{ } 3)$ and $\frac{q-1}{3} \in \mathbb{Z}^+$ $(q \in \{13,16,19,25,31,37,43,...\})$, and $i=2$ in the proposed scheme, and $r=4$, $t_0=2$ and $K_0=\frac{q-1}{3}$ in the WCCWC scheme, schemes are obtained with parameters specified in rows numbered $9$ and $10$ in Table \ref{tab:MADCC_tdes2}, respectively. In this case, it can be verified that $\frac{{\text{Load per user}}_{proposed}}{{\text{Load per user}}_{WCCWC}}=\frac{2(q^2-5q+6)}{3q(q-7)} < 1$, $\forall q$. Also, $\frac{F_{proposed}}{F_{WCCWC}}=\frac{18q}{(q-1)(q-4)} <  1$, $\forall q \ge 25$. Thus, the proposed scheme in Theorem \ref{thm:MADCC_tdes} outperforms the WCCWC scheme in terms of load per user, while the subpacketization level is lesser in all cases except when $q \in \{13,16,19\}$. To compare the proposed scheme in Theorem \ref{thm:MADCC_tdes} with the WCCWC scheme for $i > 2$, we require the existence of $t-(v,k,\lambda)$-designs with $t\ge 4$ and $\frac{v-i}{k-1} \in \mathbb{Z}^+ \backslash [i]$, which is generally difficult to satisfy. 
  	
  	The main limitation of the WCCWC scheme is that the number of users in the MADCC network is the same as the number of cache nodes. The $t$-design access topology in the proposed MADCC scheme overcomes this, where the number of users can scale linearly, polynomially or exponentially with the number of caches. We illustrate this with examples. A $t$-design in which the number of blocks is the same as the number of points is called a symmetric design. Existence of symmetric designs is discussed in \cite{Stin} and \cite{Col}. Thus, using a symmetric design, we can obtain MADCC schemes in which the number of users scales linearly with the number of caches. Row $3$ of Table \ref{tab:MADCC_tdes2} is such a case. Consider the $3-(v,4,3)$ design which exists for all even integers  $v \ge 6$, in which case $b=\frac{v(v-1)(v-2)}{8}$. For the MADCC schemes obtained using this design, the number of users scales polynomially with the number of caches. For even $n$, taking every $\frac{n}{2}$-subset of $[n]$ results in a $\frac{n}{2}-(n,\frac{n}{2},1)$ design. For the MADCC schemes obtained using this design, the number of users $K=\binom{n}{\frac{n}{2}} \approx \frac{2^n}{\sqrt{\pi n/2}}$ scales exponentially with the number of caches.  	        
\begin{table}[!htbp]
	\centering
	\setlength{\tabcolsep}{0.15pt}
	\footnotesize
	\begin{tabular}{|c|c| c | c | c | c |}
		\hline
		\rule{0pt}{4ex}
		\makecell{Sl.\\ No.}&\makecell{Schemes and Parameters}  & \makecell{$K$} & \makecell{$(\Gamma, L,\frac{M}{N})$} & \makecell{$\frac{R}{K}$}& \makecell{$F$} \\ [4pt] 	
		\hline
		\rule{0pt}{3.5ex}
		1 & \makecell{Proposed scheme, \\ $2-(9,3,1)$ design\\ (Example $1.22$ \cite{Col}), $i=1$ } &
		$12$ &
		\multirow{2}{*}{\makecell{$(9,3,\frac{1}{9})$} }&
		$\frac{1}{3}$ & \makecell{$27$} \\
		\cline{1-3}\cline{5-6}
		\rule{0pt}{3.5ex}
		2 & \makecell{WCCWC scheme, $r=3$,  \\ $t_0=1$ and $K_0=4$} & $9$
		& & $\frac{2}{3}$ & \makecell{$36$} \\
		\hline
		\rule{0pt}{3.5ex}
		3 & \makecell{Proposed scheme, \\ $2-(7,3,1)$ design\\ (Example $1.18$ \cite{Col}), $i=1$ } &
		$7$ &
		\multirow{2}{*}{\makecell{$(7,3,\frac{1}{7})$} }&
		$\frac{2}{7}$ & \makecell{$21$} \\
		\cline{1-3}\cline{5-6}
		\rule{0pt}{3.5ex}
		4 & \makecell{WCCWC scheme, $r=3$, \\ $t_0=1$ and $K_0=3$} & $7$
		& & $\frac{4}{7}$ & \makecell{$21$} \\
		\hline
		\rule{0pt}{3.5ex}
		5 & \makecell{Proposed scheme, \\ $3-(17,5,1)$ design\\ (Theorem $9.27$ \cite{Stin}), $i=1$ } &
		$68$ &
		\multirow{2}{*}{\makecell{$(17,5,\frac{1}{17})$} }&
		$\frac{4}{17}$ & \makecell{$170$} \\
		\cline{1-3}\cline{5-6}
		\rule{0pt}{3.5ex}
		6 & \makecell{WCCWC scheme, $r=5$, \\ $t_0=1$ and $K_0=4$} & $17$
		& & $\frac{12}{17}$ & \makecell{$68$} \\
		\hline
		\rule{0pt}{3.5ex}
		7 & \makecell{Proposed scheme, \\ $3-(6k+2,4,3)$ design,\\ $\forall k \ge 2$, $i=2$ } &
		$\frac{3}{4}\binom{6k+2}{3}$ &
		\multirow{2}{*}{\makecell{$(6k+2,$\\ $4,\frac{1}{3k+1})$} }&
		$\frac{3k-1}{3(3k+1)}$ & \makecell{$4(18k^2+$ \\ $9k+1)$} \\
		\cline{1-3}\cline{5-6}
		\rule{0pt}{3.5ex}
		8 & \makecell{WCCWC scheme, \\ $r=k \ge 2$, $t_0=2$ \\and $K_0=2k$} & $6k+2$
		& & $\frac{2k+1}{2(3k+1)}$ & \makecell{$4k(6k^2-$ \\ $k-1)$} \\
		\hline
		\rule{0pt}{3.5ex}
		9 & \makecell{Proposed scheme \\ $3-(q+1,4,2)$ design, \\ $q \equiv 1 (\mod 3)$,\\ $\frac{q-1}{3} \in \mathbf{Z}^+$, $i=2$} &
		$\frac{q(q^2-1)}{12}$ &
		\multirow{2}{*}{\makecell{$(q+1,$\\ $4,\frac{2}{q+1})$}} &
		$\frac{q^2-5q+6}{3q(q+1)}$ & $2q(q+1)$ \\
		\cline{1-3}\cline{5-6}
		\rule{0pt}{3.5ex}
		10 & \makecell{WCCWC scheme \\ $r=4$, $t_0=2$ \\and $K_0=\frac{q-1}{3}$} & $q+1$
		& & $\frac{q-7}{2(q+1)}$ & \makecell{$\frac{1}{9}(q^3-4q^2$ \\ $-q+4)$}  \\
		\hline
	\end{tabular}
	\caption{Comparison of the proposed scheme in Theorem \ref{thm:MADCC_tdes} and the WCCWC scheme in \cite{WCCWC}.}
	\label{tab:MADCC_tdes2}
\end{table}
\section{MADCC schemes with $t$-GDD topology}\label{madcc_schemes_tgdd}
In this section, we propose a novel MADCC scheme in which the access topology is described by a $t-(m,q,k,1)$ GDD, and obtain the derived CWEC scheme with the same topology. The proposed scheme is then compared with the derived CWEC scheme and the WCCWC scheme. Similar to Section \ref{madcc_schemes_tgdd} for the $ t$-design-based scheme,  we represent the caches and the users in the network by the points and the blocks of a given $t$-GDD, respectively, and exploit the properties of $t$-GDD in constructing the placement array and the delivery array. We first define the term \textit{$t$-GDD topology}.
\begin{defn}[$t$-GDD topology]\cite{MACC_des}
	Let $(\mathcal{X}, \mathcal{G}, \mathcal{A})$ be a $t-(m,q,k,1)$ GDD. A $t$-GDD topology is defined as an access topology $\mathfrak{B}$ in which \\
	\textit{1}. the users are represented by the blocks $A_i \in \mathcal{A}$, \\
	\textit{2}. the cache nodes are represented by the points $x \in \mathcal{X}$, \\
	\textit{3}. and the user $\{U_{A_i} : A_i \in \mathcal{A}\}$ can access all the caches $x$ such that $x \in A_i$, i.e., $\mathfrak{B}=\mathcal{A}$.
\end{defn}

\subsection{MADCC schemes with $t$-GDD topology}
In this subsection, we first propose a novel $(k,\frac{ \binom{m}{t}q^t}{\binom{k}{t}},mq,M,N)$ MADCC scheme with $t$-GDD topology given by Theorem \ref{thm:MADCC_tgdd}. We then obtain the derived CWEC scheme with $t$-GDD topology. As in Section \ref{madcc_schemes_tdes}, the MADCC schemes are characterized by the placement and delivery arrays.
\begin{thm}\label{thm:MADCC_tgdd}
	Given a $t-(m,q,k,1)$ GDD $(\mathcal{X}, \mathcal{G}, \mathcal{A})$ and if there exists an $s-(q,m,1)$ OA $(\mathcal{Y}, \mathbf{D})$ with $1 \le t \le k \le s < m$, then there exists a $ q^s\binom{k}{l}$-division $(k,\frac{ \binom{m}{t}q^t}{\binom{k}{t}},mq,M,N)$ multiaccess D2D coded caching system with access topology $\mathfrak{B}=\mathcal{A}$, for which the following memory-load point is achievable,
	\begin{equation*}
		\left(\frac{M}{N},R\right)=\left(\frac{1}{q}, R \le \frac{(q^m-q^s) q^{t-l} \binom{m-l}{t-l}}{q^s\binom{k}{l}\binom{k-l}{t-l}}\right),
	\end{equation*} for each $l \in [\min({m-s},{t-1})]$. Further, the exact value of $R$ can be computed in the following cases:
	\begin{itemize}
		\item when $k=t$ and $s=m-1$, we have scheme for $l=1$ with $R= \frac{(q-1)^t\binom{m-1}{t-1}}{t}$.
		\item when $k=t$ and $s=m-t+1$ with $s>t$, for $l=m-s$, we have scheme with $R=\frac{(q^{t-1}-1)(q-1)(m-t+1)}{t}$.
	\end{itemize}
\end{thm}
\begin{IEEEproof}
	Given a $t-(m,q,k,1)$ GDD $(\mathcal{X}, \mathcal{G}, \mathcal{A})$ and an $s-(q,m,1)$ OA $(\mathcal{Y}, \mathbf{D})$ with $1 \le t \le k \le s < m$, the cache nodes are represented by the points $x \in \mathcal{X}$, therefore, $\Gamma=mq$. The users are represented by the blocks $A_i \in \mathcal{A}$, therefore, $K=\frac{ \binom{m}{t}q^t}{\binom{k}{t}}$. The access topology $\mathfrak{B}=\mathcal{A}$, i.e., the user $\{U_{A_i} : A_i \in \mathcal{A}\}$ can access all the caches $\{x : x \in A_i\}$. Therefore, $L=|A_i|=k$. We use the point representation of GDD. That is, the $t-(m,q,k,1)$ GDD $(\mathcal{X}, \mathcal{G}, \mathcal{A})$ can be represented as follows. $\mathcal{X}=\{(u,v) : u \in [m], v \in [q]\}$, $\mathcal{G}=\{G_1,G_2,...,G_m\}$ where $G_u=\{(u,1), (u,2),...,(u,q)\}, \forall u \in [m]$ and $\mathcal{A}=\{A_1,A_2,..,A_K\}$ where $A_i =\{(u_{i,1},v_{i,1}),(u_{i,2},v_{i,2}),...,(u_{i,k},v_{i,k})\}$, $1 \le u_{i,1} <u_{i,2} < ... < u_{i,k} \le m$, $1 \le v_{i,1},v_{i,2},...,v_{i,k} \le q$. Now we describe our MADCC scheme by constructing the placement array and the delivery array. \\ \\
	\textit{I. Placement array}
	
	The cache placement is based on the $s-(q,m,1)$ OA $(\mathcal{Y}, \mathbf{D})$ where $|\mathcal{Y}|=q$ and $\mathbf{D}=\left(D_{(j,u)} \right)_{j \in [ q^s], u \in [m]}$. The $j^{th}$ row of $\mathbf{D}$ is denoted by $\vec{D}_j$. The central server first splits each file $\{W_n : n \in [N]\}$ into $ q^s$ non-overlapping packets and each packet is denoted by the rows of $\mathbf{D}$, i.e., $\{W_n=W_{n,\vec{D}_j} : j \in [ q^s], n \in [N]\}$. We further divide each packet $W_{n,\vec{D}_j}$ into  $\binom{k}{l}$ sub-packets, i.e., $\{W_{n,\vec{D}_j}=W_{n,\vec{D}_j}^{\mathcal{T}} : \mathcal{T} \in \binom{[k]}{l}, l \in [m-s], j \in [ q^s], n \in [N]\}$. Therefore, $F= q^s\binom{k}{l}$. Now we construct a  $ q^s\binom{k}{l} \times mq$ placement array $\mathbf{Z}=\left(Z_{(\vec{D}_j,\mathcal{T}),(u,v)}\right)_{j \in [ q^s], \mathcal{T} \in \binom{[k]}{l}, u \in [m], v \in [q] }$, whose rows are indexed by $\left(\vec{D}_j,\mathcal{T}\right)$ and columns are indexed by $(u,v)$, as follows.
	\begin{equation}\label{eq:pa_tgdd}
		Z_{(\vec{D}_j,\mathcal{T}),(u,v)}= \begin{cases}  \star & \text {if } D_{(j,u)}=v   \\ null & \text {otherwise. }\end{cases}
	\end{equation}
	From (\ref{eq:pa_tgdd}), it can be seen that the sub-packets cached in each cache node denoted by $(u,v) :  u \in [m], v \in [q]$ is $\mathcal{Z}_{(u,v)}=\{W_{n,\vec{D}_j}^{\mathcal{T}} : D_{(j,u)}=v, j \in [ q^s], \mathcal{T} \in \binom{[k]}{l}, n \in [N]\}\}$. Therefore, the number of sub-packets of each file cached in each cache node, which is the same as the number of $\star$s in each column of $\mathbf{Z}$, is $Z'= q^{s-1}\binom{k}{l}$. Therefore, the cache memory ratio $\frac{M}{N}=\frac{Z'}{F}=\frac{ q^{s-1}\binom{k}{l}}{ q^s\binom{k}{l}}=\frac{1}{q}$. \\ 
	
	\textit{II. Delivery array}

    A user $U_{A_i}$, where $A_i =\{(u_{i,1},v_{i,1}),(u_{i,2},v_{i,2}),...,(u_{i,k},\\ v_{i,k})\}$, can access the contents in the cache nodes $(u_{i,j},v_{i,j}), j \in [k]$. Let $\vec{\psi}(A_i)$ and $\vec{\varphi}(A_i)$ represent the vectors consisting of the first and second coordinates of each point in $A_i$, respectively, i.e., $\vec{\psi}(A_i)=\left(u_{i,1},u_{i,2},...,u_{i,k}\right)$ and  $\vec{\varphi}(A_i)=\left(v_{i,1},v_{i,2},...,v_{i,k}\right)$. Then from (\ref{eq:pa_tgdd}), the sub-packets that the user $U_{A_i}$ can retrieve, denoted by $\mathcal{Z}_{A_i}$, is $\{W_{n,\vec{D}_j}^{\mathcal{T}} : d\left(\vec{D}_j(\vec{\psi}(A_i)),\vec{\varphi}(A_i)\right) < k, j \in [ q^s], \mathcal{T} \in \binom{[k]}{l}, n \in [N]\}$, where $\vec{D}_j(\vec{\psi}(A_i))$ denotes the vector containing the entries of $\vec{D}_j$ at the indices specified in $\vec{\psi}(A_i)$. Since the number of users is $\frac{ \binom{m}{t}q^t}{\binom{k}{t}}$, we construct a  $ q^s\binom{k}{l} \times \frac{ \binom{m}{t}q^t}{\binom{k}{t}}$ delivery array $\mathbf{Q}=\left(Q_{(\vec{D}_j,\mathcal{T}),A_i} \right)_{j \in [ q^s], \mathcal{T} \in \binom{[k]}{l}, A_i \in \mathcal{A} }$, as follows.
    \begin{equation}\label{eq:da_tgdd}
    	Q_{(\vec{D}_j,\mathcal{T}),A_i}= \begin{cases}  \star & \text {if } d\left(\vec{D}_j(\vec{\psi}(A_i)),\vec{\varphi}(A_i)\right) < k   \\ \vec{e}_{\alpha(\vec{e})} & \text {otherwise, }\end{cases}
    \end{equation}
    where $\vec{e}=\left(e_1,e_2,...,e_m\right) \in [q]^m$ such that for each $n \in [m]$,
    \begin{equation}\label{eq:da_tgdd2}
    	e_n = \begin{cases}  \vec{\varphi}(A_i)(h) & \text {if } n=\vec{\psi}(A_i)(h) \in \mathcal{T}  \text{ for some } h \in [k]  \\ \vec{D}_j(n) & \text {otherwise, }\end{cases}
    \end{equation}
    and $\alpha(\vec{e})$ is the occurrence number of $\vec{e}$ in $\mathbf{Q}$ from top to bottom and left to right for a given $\vec{D}_j$.
\begin{claim}
	\label{claim:dpda_tgdd}
	The delivery array $\mathbf{Q}$ is in fact a $\left(\frac{ \binom{m}{t}q^t}{\binom{k}{t}},  q^s\binom{k}{l}, \left( q^s- (q-1)^k q^{s-k}\right)\binom{k}{l}, S \right)$ DPDA, where $S \le \frac{(q^m-q^s) q^{t-l} \binom{m-l}{t-l}}{\binom{k-l}{t-l}}$.
\end{claim} 
Proof of Claim \ref{claim:dpda_tgdd} is given in \textit{Appendix \ref{appendix:dpda_tgdd}}. The exact value of $S$ is computed in \textit{Appendix \ref{appendix:S_tgdd}} for two cases: $(i)$ when $k=t$ and $s=m-1$, and $(ii)$ when $k=t$ and $s=m-t+1$. Since $\mathbf{Q}$ is a DPDA, each user's demand can be satisfied with a load $R=\frac{S}{F} \le \frac{(q^m-q^s) q^{t-l} \binom{m-l}{t-l}}{q^s\binom{k}{l}\binom{k-l}{t-l}}$.
\end{IEEEproof}
\begin{figure}[!htbp]\centering
	\captionsetup{justification=centering}
	\includegraphics[width=0.4\textwidth]{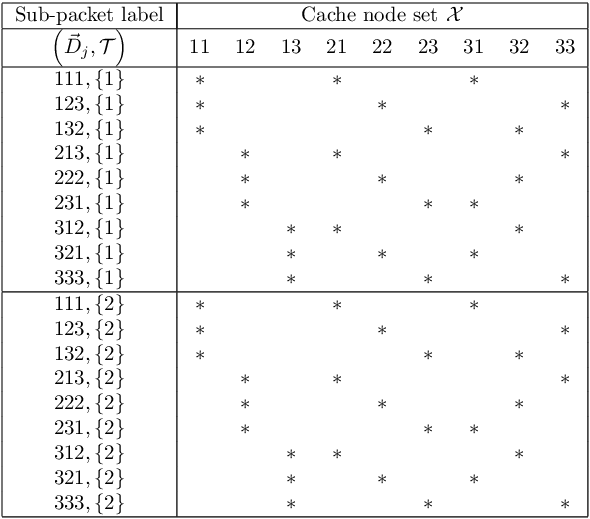}
	\caption{Placement array in Example \ref{ex:madcc_tgdd}.}
	\label{fig:pa_tgdd}
\end{figure}
\begin{figure*}[!htbp]
	\centering
	\captionsetup{justification=centering}
	\includegraphics[width=\textwidth]{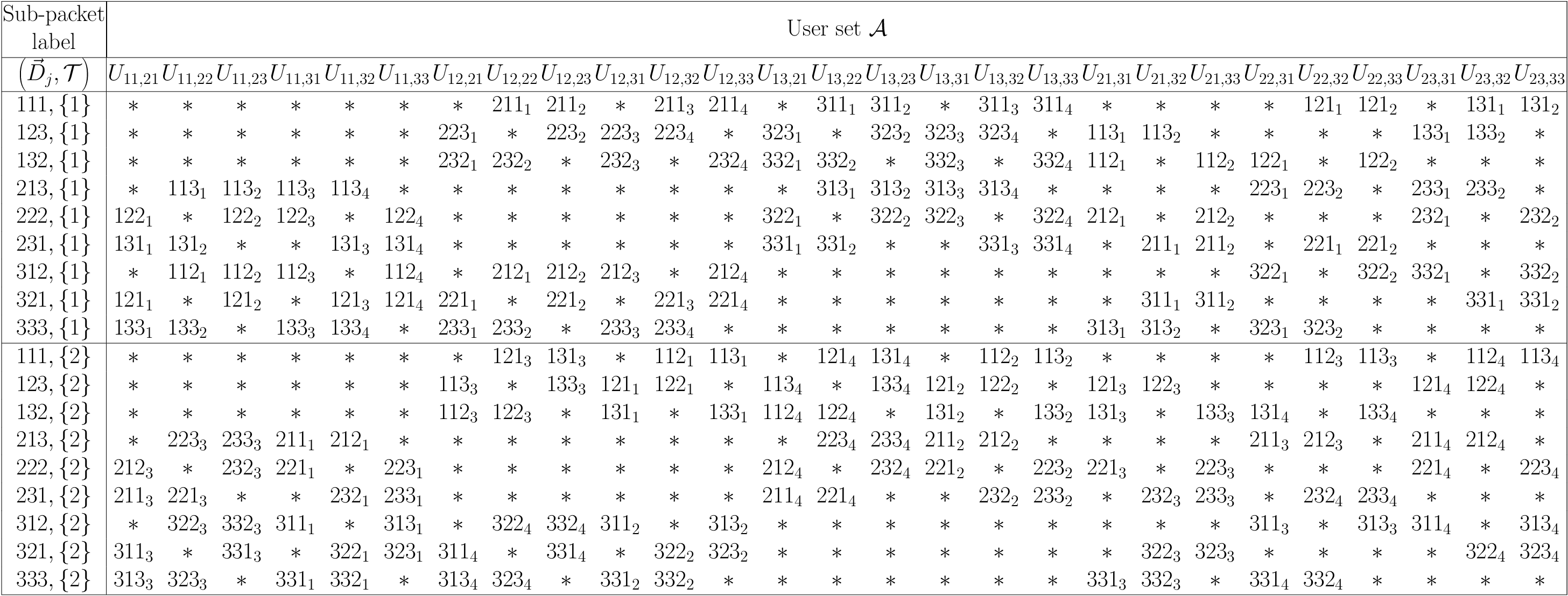}
	\caption{Delivery array in Example \ref{ex:madcc_tgdd}.}
	\label{fig:da_tgdd}
\end{figure*}

The following example illustrates Theorem \ref{thm:MADCC_tgdd}.
\begin{example}\label{ex:madcc_tgdd}
	In this example, for convenience we write a tuple $(a,b)$ as $ab$. $\mathcal{X}=\{ 11,12,13,21,22,23, 31,32,33 \}$, $\mathcal{G}=\{G_1=\{11, 12,13\}, G_2=\{21,22,23\}, G_3=\{31, 32,33\} \} $ and  $\mathcal{A}=\{\{11,21\}, \{11,22\}, \{11,23\}, \{11,31\},\{11,32\}, \\  \{11, 33\}, \{12,21\}, \{12,22\}, \{12,23\}, \{12,31\}, \{12,32\},  \{12,\\33\}, \{13,21\}, \{13,22\}, \{13,23\}, \{13,31\}, \{13,32\},  \{13,33\}, \\  \{21, 31\}, \{21,32\}, \{21,33\}, \{22,31\}, \{22,32\}, \{22,33\}, \\ \{23,31\}, \{23,32\}, \{23,33\}\}$ is a $2-(3,3,2,1)$ GDD. 	
	\begin{align*}
		\mathcal{Y}=\{1,2,3\}, \hspace{0.2cm} \mathbf{D}={\small\left(\begin{array}{ccc}
			1&1&1\\
			1&2&3\\
			1&3&2\\
			2&1&3\\
			2&2&2\\
			2&3&1\\
			3&1&2\\
			3&3&3
		\end{array}\right) }\text{ is a $2-(3,3,1)$ OA}.
	\end{align*} 

	 By  using this $2-(3,3,2,1)$ GDD and $2-(3,3,1)$ OA, for $l=1$, a $18$-division $(2,27,9,3,9)$ MADCC scheme with access topology $\mathfrak{B}=\mathcal{A}$ can be obtained as follows. The $9$ cache nodes are represented by the $9$ points in $\mathcal{X}$, and the $27$ users are represented by the $27$ blocks in $\mathcal{A}$, and each user can access $2$ cache nodes that are represented by the $2$ points in the corresponding block. Each file is divided into $9 \times 2 =18$ sub-packets, i.e., $\{W_{n}=W_{n,\vec{D}_j}^{\mathcal{T}} : \mathcal{T} \in \binom{[2]}{1}, j \in [ 9], \forall n \in [18]\}$. Then by (\ref{eq:pa_tgdd}), we obtain the $18 \times 9$ placement array shown in Fig.\ref{fig:pa_tgdd}.
			
By (\ref{eq:da_tgdd}), we obtain the $18 \times 27$ delivery array shown in Fig.\ref{fig:da_tgdd}. 

To make clear how the subscript $\alpha(\vec{e})$ is written, consider the case $\vec{D}_j=111$ and  $\vec{e}=131$. The vectors $\vec{D}_j$ and $\vec{e}$ differ at the second index, where $\vec{e}(2)=3$. Therefore, by (\ref{eq:da_tgdd2}), $\vec{e}=131$ will appear in those $A_i$ in which $d\left(111(\vec{\psi}(A_i)),\vec{\varphi}(A_i)\right)=2$ and $(2,\vec{e}(2))=(2,3) \subset A_i$. There are $4$ such columns. Therefore, $\vec{e}=131$ appears $4$ times all together in the rows $\left(111,\mathcal{T}\right)$, where $\mathcal{T} \in \binom{[2]}{1}$. Therefore, $\alpha(131)$ takes values from $1$ to $4$ depending on the occurrence of $131$ from top to bottom and left to right in $\mathbf{Q}$, in the rows corresponding to $\vec{D}_j=111$. Note that $\alpha(\vec{e})$ takes values from $1$ to $4$ for all $\vec{e}$.  

It can be verified that the delivery array in Fig.\ref{fig:da_tgdd} is a $(27,18,10,72)$ DPDA and therefore, using Algorithm \ref{d2d_alg}, we obtain the scheme with load $R=\frac{S}{F}=4$ and subpacketization level $F=18$. The mapping $\phi$ from the non-$\star$ entries $\vec{e}_{\alpha(\vec{e})}$  to the users $\{U_{A_{i'}} : A_{i'} \in \mathcal{A}\}$, $\{\phi\left(\vec{e}_{\alpha(\vec{e})}\right)= U_{A_{i'}}\}$, is such that the block $U_{A_{i'}}$ contains any of the $t$-set of points among the points $\{(1,e_1),(2,e_2),...,(m,e_m)\}$. For example, consider the non-$\star$ entry $113_1$. The user which performs the coded transmission corresponding to the non-$\star$ entry $113_1$ is the user $U_{A_{i'}}$ such that the block $A_{i'}$ contains any of the $2$-set of points among the points $\{(1,1),(2,1),(3,3)\}$. There are exactly $3$ users, $U_{11,21}, U_{11,33}$ and $U_{21,33}$, satisfying this condition and anyone can perform the transmission.  	
\end{example}
\begin{rem}
	Given a $t-(m,q,k,1)$ GDD $(\mathcal{X}, \mathcal{G}, \mathcal{A})$, one can always obtain an MADCC scheme for a $(k,\frac{ \binom{m}{t}q^t}{\binom{k}{t}},mq,M,N)$ MADCC network for $\frac{M}{N}=\frac{1}{q}$ from Theorem \ref{thm:MADCC_tgdd} using the $(m-1)-(q,m,1)$ OA specified in Lemma \ref{triv_OA}. The availability of other $s-(q,m,1)$ OAs and other choices of $l$ leads to schemes with different load and subpacketization level for the same MADCC system. 
\end{rem}
Next, we discuss the derived CWEC scheme with $t$-GDD topology. We consider a case where the user-retrieve array (equivalent to the delivery array) designed for the MACC scheme with $t$-GDD topology in \cite{MACC_des} is a $(g+1)-(K, F, Z, S)$ regular PDA and the exact value of $S$ is known. The placement array of the derived CWEC scheme with $t$-GDD topology will be the same as the node-placement array designed for the MACC scheme with $t$-GDD topology in \cite{MACC_des}, except that the number of sub-packets to which each packet is divided gets increased by a factor of $g$. For a given $t-(m,q,k,1)$ GDD and a $s-(q,m,1)$ OA with $1 \le t \le k \le s \le m$, the user-retrieve array of the MACC scheme with $t$-GDD topology in  \cite{MACC_des} is a $\left(\frac{ \binom{m}{t}q^t}{\binom{k}{t}}, q^s\binom{k}{t}, \left(q^s-(q-1)^kq^{s-k}\right)\binom{k}{t}, S \right)$ PDA, where $S \le (q-1)^tq^m$. When $k=t$ and $s=m-1$, $S=(q-1)^tq^{m-1}$ and the regularity of PDA is $g=\binom{m}{t}$. Therefore, when $k=t$ and $s=m-1$, the delivery array of the derived CWEC scheme with $t$-GDD topology is a  $\left(\binom{m}{t}q^t, \left(\binom{m}{t}-1\right)q^{m-1}, \left(\binom{m}{t}-1\right)q^{m-1}\left(1-\left(\frac{q-1}{q}\right)^t\right), \right. \\ \left. \binom{m}{t}(q-1)^tq^{m-1} \right)$ DPDA. In the resulting MADCC scheme, there are $mq$ caches each having a memory $\frac{M}{N}=\frac{1}{q}$. The transmission load is given by, $R=\frac{S}{F}=\frac{\binom{m}{t}(q-1)^t}{\binom{m}{t}-1}$. This scheme is stated as the following lemma.
\begin{table*}[!hbp]
	\begin{center}
		\setlength{\tabcolsep}{4.2pt}
		\begin{tabular}{|c|c|c|c|c|}
			\hline
			\textbf{\makecell{Schemes and parameters}}  & \textbf{$K$} & \textbf{$R$} & \textbf{$\frac{R}{K}$} & \textbf{ $F$} \\\hline
			\makecell{Proposed scheme with $t$-GDD topology in Theorem \ref{thm:MADCC_tgdd},\\ $t-(m,q,t,1)$ GDD,  $(m-1)-(q,m,1)$ OA and $l=1$}
			& $\binom{m}{t}q^t$ & $\frac{\binom{m-1}{t-1}(q-1)^t}{t}$ & $\frac{1}{m}\left(\frac{q-1}{q}\right)^t$ & $t q^{m-1}$ \\\hline
			\makecell{Derived CWEC scheme with \\ $t$-GDD topology in Lemma \ref{lem:derived_MADCC_tgdd}, \\ $t-(m,q,t,1)$ GDD and $(m-1)-(q,m,1)$ OA} 
			& $\binom{m}{t}q^t$  & $\frac{\binom{m}{t}(q-1)^t}{\binom{m}{t}-1}$ & $\frac{1}{\binom{m}{t}-1}\left(\frac{q-1}{q}\right)^t$ & $\left(\binom{m}{t}-1\right)q^{m-1}$ \\\hline
			\makecell{WCCWC scheme in \cite{WCCWC}, integer $r=t>1$,\\ MAN PDA with $t_0=m$ and $K_0=\frac{m(q-1)}{r-1} \in \mathbb{Z}^+ \backslash [m-1]$}
			& $mq$ & $q-t$ & $\frac{q-t}{mq}$ & $\binom{\frac{m(q-1)}{t-1}}{m}m^2q$ \\\hline
		\end{tabular}
	\end{center}
	\caption{Comparison between the proposed scheme in Theorem \ref{thm:MADCC_tgdd}, the derived CWEC scheme in Lemma \ref{lem:derived_MADCC_tgdd} and the WCCWC scheme, for $\Gamma=mq$, $L=t$ and $\frac{M}{N}=\frac{1}{q}$.}
	\label{tab:MADCC_tgdd}
\end{table*}
\begin{lem}\label{lem:derived_MADCC_tgdd}
	(Derived CWEC scheme with $t$-GDD topology) Given a $t-(m,q,t,1)$ GDD $(\mathcal{X}, \mathcal{G}, \mathcal{A})$ and an $(m-1)-(q,m,1)$ OA $(\mathcal{Y}, \mathbf{D})$, there exists an $ \left(\binom{m}{t}-1\right)q^{m-1}$-division $(t,\binom{m}{t}q^t,mq,M,N)$ multiaccess D2D coded caching system with access topology $\mathfrak{B}=\mathcal{A}$, for which the following memory-load point is achievable,
	\begin{equation*}
		\left(\frac{M}{N},R\right)=\left(\frac{1}{q}, \frac{\binom{m}{t}(q-1)^t}{\binom{m}{t}-1}\right).
	\end{equation*}
\end{lem}
Note that, for any positive integers $t$, $m$ and $q$, with $1<t<m$ and $q \geq 2$, there exists an $(m-1)-(q,m,1)$ OA given by Lemma \ref{triv_OA} and a $t-(m,q,t,1)$ GDD given by Lemma \ref{triv_GDD}. 
\subsection{Performance Analysis of the proposed MADCC scheme with $t$-GDD topology}
In this subsection, we carry out a comparison of the proposed MADCC scheme with $t$-GDD topology in Theorem \ref{thm:MADCC_tgdd}, the derived CWEC scheme with $t$-GDD topology in Lemma \ref{lem:derived_MADCC_tgdd} and the WCCWC scheme, using the metrics of load per user and subpacketization level while keeping the number of caches, cache size and access degree the same. By choosing the parameters, a $t-(m,q,t,1)$ GDD, an $(m-1)-(q,m,1)$ OA and $l=1$  for the proposed scheme in Theorem \ref{thm:MADCC_tgdd}, a $t-(m,q,t,1)$ GDD and an $(m-1)-(q,m,1)$ OA for the derived CWEC scheme in Lemma \ref{lem:derived_MADCC_tgdd}, for positive integers $r=t>1$, $t_0=m$ and $K_0=\frac{m(q-1)}{r-1} \in \mathbb{Z}^+ \backslash [m-1]$ for the WCCWC scheme, MADCC schemes are obtained with $mq$ number of caches, memory ratio $\frac{M}{N}$=$\frac{1}{q}$ and access degree $L=t$. These schemes are summarized in Table \ref{tab:MADCC_tgdd}. 

From Table \ref{tab:MADCC_tgdd}, it is seen that under the same $t$-GDD topology, $\frac{F_{proposed}}{F_{derived}}=\frac{t}{\binom{m}{t}-1}$ and $\frac{{\text{Load per user}}_{proposed}}{{\text{Load per user}}_{derived}}=\frac{\binom{m}{t}-1}{m}$. That is, when $t=m-1$, for any $q$, both the proposed scheme in Theorem \ref{thm:MADCC_tgdd} and the derived CWEC scheme in Lemma \ref{lem:derived_MADCC_tgdd} have the same subpacketization level, but the load per user of the proposed scheme is $\frac{m-1}{m}$ of that of the derived CWEC scheme with $t$-GDD topology. For other values of $m$ and $t$, the proposed MADCC scheme with $t$-GDD topology requires a lesser subpacketization level at the expense of an increase in load per user. Comparing the proposed MADCC scheme in Theorem \ref{thm:MADCC_tgdd} with the WCCWC scheme, it is seen that the proposed scheme has a significant advantage in terms of subpacketization level at the expense of an increase in load per user, for a given $\Gamma$, $\frac{M}{N}$, and $L$. The MADCC scheme with $t$-GDD topology supports more users than the WCCWC scheme. The following example illustrates the above comparisons.
 \begin{example}
From a $2-(3,3,2,1)$ GDD and a $2-(3,3,1)$ OA, Theorem \ref{thm:MADCC_tgdd} gives a $18$-division $(2,27,9,M,N)$ MADCC scheme, detailed in Example \ref{ex:madcc_tgdd}, with  $\frac{M}{N}=\frac{1}{3}$, load $R=4$ and load per user $\frac{R}{K}=0.148$. Using the same $2-(3,3,2,1)$ GDD and $2-(3,3,1)$ OA, the derived CWEC scheme in Lemma \ref{lem:derived_MADCC_tgdd} gives a $18$-division $(2,27,9,M,N)$ MADCC scheme with cache memory $\frac{M}{N}=\frac{1}{3}$, load $R=6$ and load per user $\frac{R}{K}=0.222$. For $r=2$, and the MAN PDA with $t_0=3$ and $K_0=6$, WCCWC scheme gives a $540$-division $(2,9,9,M,N)$ MADCC scheme with cache memory $\frac{M}{N}=\frac{1}{3}$, load $R=1$ and load per user $\frac{R}{K}=0.111$. That is, for the same number of caches, cache memory, and access degree, the proposed MADCC scheme in Theorem \ref{thm:MADCC_tgdd} requires a subpacketization level of $18$, whereas the scheme in \cite{WCCWC} with cyclic wrap-around topology requires a significantly higher subpacketization level of $540$, though with a small increase in load per user from $0.111$ to $0.148$.
 \end{example}
\section{D2D Coded Caching Schemes for the original D2D network}\label{d2d_madcc}
In this section, we obtain coded caching schemes for the original D2D network with dedicated caches.
Since the delivery arrays of the MADCC schemes discussed in Section \ref{madcc_schemes_tdes} and Section \ref{madcc_schemes_tgdd}  are DPDAs, coded caching schemes for the original D2D network can be obtained directly from the delivery arrays of the proposed MADCC schemes by using Algorithm \ref{d2d_alg}, as stated in the following corollaries.
\begin{cor}\label{cor:dpda1}
	(DPDA via Theorem \ref{thm:MADCC_tdes}) Given a $t-(v,k,\lambda)$ design, for any $i \in [t-1]$, there exists a $\left(\frac{\lambda \binom{v}{t}}{\binom{k}{t}},  \binom{v}{i}\binom{k}{t-i},  \left(\binom{v}{i}-\binom{v-k}{i}\right)\binom{k}{t-i},  \frac{\lambda\binom{v}{t}\binom{v-t}{k-t+i}}{\binom{v-t}{k-t}} \right)$ DPDA which gives a D2D coded caching scheme with memory ratio $\frac{M}{N}=1-\frac{\binom{v-k}{i}}{\binom{v}{i}}$ and load $R=\frac{\lambda\binom{v-i}{k}}{\binom{v-t}{k-t}\binom{t}{i}}$. 
\end{cor}
\begin{cor}\label{cor:dpda2}
	(DPDA via Corollary \ref{cor:MADCC_tdes}) Given a $t-(v,k,\lambda)$ design, for any $i \in [\min(t,k-t+1)]$, there exists a $\left(\frac{\lambda \binom{v}{t}}{\binom{k}{t}}, \binom{v}{i}\binom{k}{t-1}, \left(\binom{v}{i}-\binom{v-k}{i}\right)\binom{k}{t-1}, \frac{\lambda\binom{v}{i+t-1}\binom{v-(i+t-1)}{k-t+1}}{\binom{v-t}{k-t}} \right)$ DPDA which gives a D2D coded caching scheme with memory ratio $\frac{M}{N}=1-\frac{\binom{v-k}{i}}{\binom{v}{i}}$ and transmission load $R=\frac{\lambda\binom{v-i}{k}}{\binom{v-t}{k-t}\binom{i+t-1}{i}}$, provided for any $\mathcal{Y} \subseteq A \in \mathcal{A}$ with $t < |\mathcal{Y}| \le k$, the number of blocks in $\mathcal{A}$ that contain $\mathcal{Y}$, $\lambda(\mathcal{Y})$, is same for all $\mathcal{Y}$. 
\end{cor} 
\begin{cor}\label{cor:dpda_tgdd}
	(DPDA via Theorem \ref{thm:MADCC_tgdd}) Given a $t-(m,q,k,1)$ GDD $(\mathcal{X}, \mathcal{G}, \mathcal{A})$ and if there exists a $s-(q,m,1)$ OA $(\mathcal{Y}, \mathbf{D})$ with $1 \le t \le k \le s < m$, then there exists an $\left(\frac{ \binom{m}{t}q^t}{\binom{k}{t}},  q^s\binom{k}{l}, \left( q^s- (q-1)^k q^{s-k}\right)\binom{k}{l}, S \le \frac{(q^m-q^s) q^{t-l} \binom{m-l}{t-l}}{\binom{k-l}{t-l}} \right)$ DPDA, for each $l \in [\min({m-s},{t-1})]$, which gives a D2D coded caching scheme with memory ratio $\frac{M}{N}=1-\left(\frac{q-1}{q}\right)^k$ and transmission load $R \le \frac{(q^m-q^s) q^{t-l} \binom{m-l}{t-l}}{q^s\binom{k}{l}\binom{k-l}{t-l}}$.  Further, the exact value of $R$ can be computed in the following cases:
	\begin{itemize}
		\item when $k=t$ and $s=m-1$, we have scheme for $l=1$ with $R= \frac{(q-1)^t\binom{m-1}{t-1}}{t}$.
		\item when $k=t$ and $s=m-t+1$ with $s>t$, for $l=m-s$, we have scheme with $R=\frac{(q^{t-1}-1)(q-1)(m-t+1)}{t}$.
	\end{itemize}
\end{cor} 

From Corollary \ref{cor:dpda1}, we can obtain the following corollary.
\begin{cor}\label{cor:dpda3}
	For any positive integers $k$ and $n$ such that $k \le \frac{n}{2}$, there exists an $(\binom{n}{k},M,N)$ D2D coded caching scheme, for which the following memory-load points are achievable,
	\begin{equation*}
		\begin{split}
	&	\left(\frac{M}{N},R\right)=\left(1-\frac{\binom{n-k}{i}}{\binom{n}{i}}, \frac{\binom{n-i}{k}}{\binom{k-1}{i}}\right) \text{ with } F=\binom{n}{i}\binom{k}{i+1}, \\ & \text{ for each } i \in [k-2], \text { and }
	\\ &
		\left(\frac{M}{N},R\right)=\left(1-\frac{\binom{k}{j}}{\binom{n}{j}}, \frac{\binom{n-j}{k-j}}{\binom{k-1}{j}}\right) \text{ with } F=\binom{n}{j}\binom{n-k}{k-j-1},
	\\ & \text{ for each }  j \in [\min(n-k-2,k-1)].
	\end{split} 
\end{equation*}
\end{cor}
Proof of Corollary \ref{cor:dpda3} is given in Appendix \ref{appendix:dpda3}.  The following example illustrates Corollary \ref{cor:dpda3}.
\begin{example}\label{ex:dpda3}
	For $n=8$ and $k=3$, by Corollary \ref{cor:dpda3}, we obtain a $(56,M,N)$ D2D scheme at $4$ memory points as in Table \ref{tab:DPDA3}. 
\end{example}	
	\begin{table}[H]
	\centering
	\begin{tabular}{| c | c | c | c | }
		\hline
		\rule{0pt}{4ex}
		$i/j$ & $\frac{M}{N}$ & $F$ &  $R$ \\ [4pt] 	
		\hline
		\rule{0pt}{3.5ex}
		$i=1$  & $0.375$ & $24$ & $17.5$\\	
		\hline
		\rule{0pt}{3.5ex}
		$j=1$ & $0.625$ & $80$ & $5.25$\\
		\hline
		\rule{0pt}{3.5ex}
		$j=2$ & $0.892$ & $280$ & $1$\\
		\hline
		\rule{0pt}{3.5ex}
		$j=3$ & $0.982$ & $280$ & $0.25$\\
		\hline
	\end{tabular}
	\centering
	\caption{D2D coded caching schemes by Corollary \ref{cor:dpda3} for $n=8, k=3$ for different $i$ and $j$ values. }
	\label{tab:DPDA3}
	\end{table}

From Corollary \ref{cor:dpda_tgdd}, we can obtain the following corollary.
\begin{cor}\label{cor:dpda4}
	For any positive integers $t$ and $m$ with $1<t<m$, and $q \geq 2$, there exists an  $(\binom{m}{t}q^t,M,N)$ D2D coded caching scheme with $F=q^{m-1}t$, $\frac{M}{N}=1-\left(\frac{q-1}{q}\right)^t$  and $R= \frac{(q-1)^t\binom{m-1}{t-1}}{t}$.
\end{cor}
\begin{IEEEproof}
	For any positive integers $t$ and $m$ with $1<t<m$, and $q \geq 2$, there always exists a $(m-1)-(q,m,1)$ OA $(\mathcal{Y}, \mathbf{D})$ given by Lemma \ref{triv_OA} and a $t-(m,q,t,1)$ GDD $(\mathcal{X}, \mathcal{G}, \mathcal{A})$ given by Lemma \ref{triv_GDD}. Using this $(m-1)-(q,m,1)$ OA and a $t-(m,q,t,1)$ GDD, Corollary \ref{cor:dpda_tgdd} gives a $\left( \binom{m}{t}q^t,  q^{m-1}t, \left( q^{m-1}- (q-1)^t q^{m-t-1}\right)t, q^{m-1} (q-1)^{t} \right. \\ \left. \binom{m-1}{t-1} \right)$ DPDA. Using this DPDA, Algorithm \ref{d2d_alg} gives a $(\binom{m}{t}q^t,M,N)$ D2D coded caching scheme with $F=q^{m-1}t$, $\frac{M}{N}=1-\left(\frac{q-1}{q}\right)^t$  and $R= \frac{(q-1)^t\binom{m-1}{t-1}}{t}$.
\end{IEEEproof}

Now we compare the transmission load and the subpacketization level of above obtained D2D schemes in Corollaries $\ref{cor:dpda1}$, $\ref{cor:dpda2}$ and $\ref{cor:dpda_tgdd}$ with that of the known D2D coded caching schemes with the same $K$ and $\frac{M}{N}$.  

\hspace{1em} $\bullet$ \textit{Comparison with the JCM scheme}: First we compare the obtained D2D schemes in Corollary \ref{cor:dpda1} and Corollary \ref{cor:dpda2} with the JCM scheme. For $K=\frac{\lambda \binom{v}{t}}{\binom{k}{t}}$ and $\frac{M}{N}=1-\frac{\binom{v-k}{i}}{\binom{v}{i}}$, we have $\frac{KM}{N}=\frac{\lambda \binom{v}{t}}{\binom{k}{t}}\left(1-\frac{\binom{v-k}{i}}{\binom{v}{i}}\right)=\frac{\lambda \binom{v}{t}}{\binom{k}{t}}-\frac{\lambda\binom{v}{t}\binom{v-k}{i}}{\binom{k}{t}\binom{v}{i}}=\frac{\lambda \binom{v}{t}}{\binom{k}{t}}-\frac{\lambda\binom{v-i}{k}}{\binom{v-t}{k-t}}=K-\lambda_0^i$, where $\lambda_0^i$ is the number of blocks that contain none of the $i$ points. Since $\frac{KM}{N} \le K$ is an integer, for the same number of users $K$ and memory ratio $\frac{M}{N}$ in Corollary \ref{cor:dpda1} and Corollary \ref{cor:dpda2}, the JCM scheme gives a transmission load $R=\frac{\binom{v-k}{i}}{\binom{v}{i}-\binom{v-k}{i}}$ and subpacketization level $F=\left(K-\lambda_0^i\right)\binom{K}{K-\lambda_0^i}$. That is, compared to the JCM scheme, proposed schemes in Corollary \ref{cor:dpda1} and Corollary \ref{cor:dpda2} have significantly lesser subpacketization level at the expense of an increase in transmission load. The following example illustrates this.
\begin{example}\label{ex:d2d_tdes1}
	For a $3-(10,4,1)$ design and $i=2$, both Corollary \ref{cor:dpda1} and  Corollary \ref{cor:dpda2} give D2D scheme at the memory point $M=20$ for $K=30$ and $N=30$, with $(R,F)=(3.333,180)$ and $(R,F)=(1.667,270)$, respectively. For the same system parameters, the JCM scheme has $R= 0.5$ and $F= 6.009 \times 10^8$.
\end{example}

For $k=t$ and $s=m-1$, Corollary \ref{cor:dpda_tgdd} give D2D schemes with $K=\binom{m}{t}q^t$, $\frac{M}{N}=1-\left(\frac{q-1}{q}\right)^t$, $F=q^{m-1}t$ and $R = \frac{(q-1)^t\binom{m-1}{t-1}}{t}$. For this $K$ and $\frac{M}{N}$, we have $\frac{KM}{N}=\binom{m}{t}(q^t-(q-1)^t)=K' < K$ is an integer and therefore, the JCM scheme requires a subpacketization level $F=K' \binom{K}{K'}$ and has $R=\frac{(q-1)^t}{q^t-(q-1)^t}$. Thus, compared to the JCM scheme, proposed schemes in Corollary \ref{cor:dpda_tgdd} have significantly lesser subpacketization level at the expense of an increase in transmission load. 

\hspace{1em} $\bullet$ \textit{Comparison with D2D schemes obtainable by transforming regular PDAs}: In the literature, there exist no direct DPDA constructions with the same $K$ and $\frac{M}{N}$ as in  Corollaries $\ref{cor:dpda1}$, $\ref{cor:dpda2}$ and $\ref{cor:dpda_tgdd}$. Given a $(g+1)-(K,F,Z,S)$ PDA, Wang \textit{et al.} in \cite{JMQX} proposed a method to construct a $(K,gF,gZ,(g+1)S)$ DPDA. That is, given a $(g+1)-(K,F,Z,S)$ PDA, we can obtain a D2D scheme with $K$ users, $\frac{M}{N}=\frac{Z}{F}$, subpacketization level $gF$ and transmission load $\frac{(g+1)S}{gF}$.  We compare our proposed DPDAs with the schemes obtained by transforming known regular PDAs with the same $K$ and $\frac{M}{N}$. Given a $t-(v,k,1)$ design and $i=1$, \cite{MACC_des} constructs a  $(t+1)-\left(\frac{ \binom{v}{t}}{\binom{k}{t}}, v\binom{k}{t}, k\binom{k}{t}, \binom{v}{t+1}-\frac{\binom{v}{t}}{\binom{k}{t}}\binom{k}{t+1} \right)$ PDA. Transforming this PDA gives a D2D scheme with $K=\frac{ \binom{v}{t}}{\binom{k}{t}}$, $\frac{M}{N}=\frac{k}{v}$, $F=tv\binom{k}{t}$ and $R= \frac{\binom{v-1}{k}}{\binom{v-t}{k-t}t}$. From a $t-(v,k,1)$ design, for $i=1$, Corollary \ref{cor:dpda1} and Corollary \ref{cor:dpda2} give a D2D scheme, for the same $K$ and $\frac{M}{N}$, with $F=v\binom{k}{t-1}$ and $R=\frac{\binom{v-1}{k}}{\binom{v-t}{k-t}t}$. That is, Corollary \ref{cor:dpda1} and Corollary \ref{cor:dpda2} give schemes that achieve the same transmission load as the scheme in \cite{MACC_des} with a subpacketization level reduced by a factor of $(k-t+1)$. 

For any positive integers $q,m$ and $t$ with $q \geq 2$ and $t <m$, \cite{CJYT} constructs a $\binom{m}{t}-\left(\binom{m}{t}q^t, q^m, q^m-q^{m-t}(q-1)^t, q^m(q-1)^t\right)$ PDA. For any positive integers $q$, $m$ and $t$ with $q \geq 2$ and $t <m$, \cite{MACC_des} and \cite{MJXQ} give a $\binom{m}{t}-\left(\binom{m}{t}q^t, q^{m-1}, q^{m-1}-q^{m-t-1}(q-1)^t, q^{m-1}(q-1)^t \right)$ PDA. That is, for the same $K=\binom{m}{t}q^t$ and $\frac{M}{N}=1-\left(\frac{q-1}{q}\right)^t$, both PDAs achieve the same $R=(q-1)^t$, but the subpacketization level of the second PDA is reduced by a factor of $q$. By transforming these two regular PDAs, D2D schemes can be obtained with $F=\left(\binom{m}{t}-1\right)q^m$, $R=\frac{\binom{m}{t}-1}{\binom{m}{t}}(q-1)^t$ and $F=\left(\binom{m}{t}-1\right)q^{m-1}$, $R=\frac{\binom{m}{t}-1}{\binom{m}{t}}(q-1)^t$, respectively. Comparing these D2D schemes with the scheme in Corollary \ref{cor:dpda4}, it can be seen that the proposed scheme in Corollary \ref{cor:dpda4} has the advantage in subpacketization level at the expense of an increase in transmission load.

In addition to the D2D schemes for the original D2D network obtained from the proposed MADCC schemes, another novel D2D scheme is obtained using $t$-GDDs and orthogonal arrays, as stated in the following corollary.
\begin{cor}\label{cor:dpda_new}
	For any positive integers $t$ and $m$ with $1<t<m$, and $q \geq 2$, there exists a  $(q^{m-1},M,N)$ D2D coded caching scheme with $F=\binom{m}{t}q^t$, $\frac{M}{N}=1-\left(\frac{q-1}{q}\right)^t$  and $R= \frac{(q-1)^tq^{m-t-1}}{\binom{m}{t}}$.
\end{cor}
\begin{IEEEproof}
	For any positive integers $t$, $m$ and $q$, with $1<t<m$ and $q \geq 2$, there always exists a $(m-1)-(q,m,1)$ OA $(\mathcal{Y}, \mathbf{D})$ given by Lemma \ref{triv_OA} and a $t-(m,q,t,1)$ GDD $(\mathcal{X}, \mathcal{G}, \mathcal{A})$ given by Lemma \ref{triv_GDD}. $\vec{\psi}(A_i)$ and $\vec{\varphi}(A_i)$ represent the vectors consisting of the first and second coordinates of each point in $A_i$, respectively, for each block $A_i \in \mathcal{A}$. We construct an array $\mathbf{P}$  whose rows are indexed by the blocks $A \in \mathcal{A}$, and columns are indexed by the rows of $\mathbf{D}$, i.e., $\{\vec{D}_j, j \in [q^{m-1}]\}$. Since the number of blocks in $\mathcal{A}$ is $\binom{m}{t}q^t$ and number of rows in $\mathbf{D}$ is $q^{m-1}$, we construct a $\binom{m}{t}q^t \times q^{m-1}$ array  $\mathbf{P}=\left(P_{A_i,\vec{D}_j} \right)_{A_i \in \mathcal{A}, j \in [ q^{m-1}] }$, as follows. 
\begin{equation}\label{eq:d2d_new}
	P_{A_i,\vec{D}_j}= \begin{cases}  \star & \text {if } d\left(\vec{D}_j(\vec{\psi}(A_i)),\vec{\varphi}(A_i)\right) < t   \\ \vec{e}_{\alpha(\vec{e})} & \text {otherwise, }\end{cases}
\end{equation}
where $\vec{e}=\left(e_1,e_2,...,e_m\right) \in [q]^m$ such that for each $i \in [m]$,
\begin{equation}\label{eq:d2d_new2}
	e_i = \begin{cases}  \vec{\varphi}(A_i)(h) & \text {if } i=\vec{\psi}(A_i)(h) \text{ for some } h \in [t]  \\ \vec{D}_j(i) & \text {otherwise, }\end{cases}
\end{equation}
and $\alpha(\vec{e})$ is the occurrence number of $\vec{e}$ from left to right in a given row $A_i$.
\begin{claim}
	\label{claim:dpda_new}
	The array $\mathbf{P}$ is in fact a $\left(q^{m-1}, \binom{m}{t}q^t, \left(q^t-(q-1)^t\right)\binom{m}{t}, (q-1)^tq^{m-1}\right)$ DPDA.
\end{claim} 
Proof of Claim \ref{claim:dpda_new} is given in \textit{Appendix \ref{appendix:dpda_new}}. By Theorem \ref{thm:d2d}, the obtained DPDA gives a $(q^{m-1},M,N)$ D2D coded caching scheme with $F=\binom{m}{t}q^t$, $\frac{M}{N}=1-\left(\frac{q-1}{q}\right)^t$  and $R= \frac{(q-1)^tq^{m-t-1}}{\binom{m}{t}}$.	
\end{IEEEproof}

\begin{figure}[!htbp]
	\centering
	\captionsetup{justification=centering}
	\includegraphics[width=0.3\textwidth]{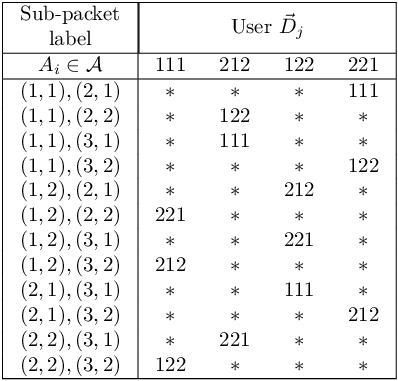}
	\caption{DPDA in Example \ref{ex:dpda_new}.}
	\label{fig:dpda_new}
\end{figure}

The following example illustrates Corollary \ref{cor:dpda_new}.
\begin{example}\label{ex:dpda_new}
	Consider $t=2$, $q=2$ and $m=3$. For $q=2$ and $m=2$, Lemma \ref{triv_OA}  gives a $2-(2,3,1)$ OA $(\mathcal{Y}, \mathbf{D})$ where, $\mathcal{Y}=\{1,2\}$ and $\mathbf{D}= \left(\begin{array}{ccc}
		1&1&1\\
		2&1&2\\
		1&2&2\\
		2&2&1
	\end{array}\right)$. For $t=2$, $q=2$ and $m=3$, Lemma \ref{triv_GDD} gives a $2-(3,2,2,1)$ GDD with $\mathcal{X}=\{(1,1),(1,2),(2,1),(2,2),(3,1),(3,2)\}$, $\mathcal{G}=\{G_1=\{(1,1), (1,2)\},G_2=\{(2,1), (2,2)\}, G_3=\{(3,1), (3,2)\} \}$  and $\mathcal{A}=\{\{(1,1),(2,1)\}, \{(1,1),(2,2)\},\{(1,1),(3,1)\}, \\  \{(1,1),(3,2)\},\{(1,2),(2,1)\},\{(1,2),(2,2)\},\{(1,2),(3,1)\}, \\  \{(1,2),(3,2)\},\{(2,1), (3,1)\},\{(2,1),(3,2)\},\{(2,2),(3,1)\}, \\ \{(2,2),(3,2)\}\}$. An array whose rows are indexed by the blocks $A \in \mathcal{A}$, and columns are indexed by the rows of $\mathbf{D}$ can be obtained using (\ref{eq:d2d_new}) and (\ref{eq:d2d_new2}), as shown in Fig.\ref{fig:dpda_new}. In this example, we discard the subscript $\alpha$ since $\alpha$ takes the value only $1$. It can be verified that the array in Fig.\ref{fig:dpda_new} is a $(4,12,9,4)$ DPDA. Therefore, by using Algorithm \ref{d2d_alg}, a $(4,M,N)$ D2D coded caching scheme can be obtained with $F=12$, $\frac{M}{N}=\frac{3}{4}$ and $R=\frac{1}{3}$.  
\end{example}

\hspace{1em} $\bullet$ \textit{Comparison of the scheme in Corollary \ref{cor:dpda_new} with the JCM scheme}:  For $K=q^{m-1}$ and $\frac{M}{N}=1-\left(\frac{q-1}{q}\right)^t$, we have $\frac{KM}{N}=q^{m-1}-q^{m-1-t}(q-1)^t$. Since $\frac{KM}{N} < K$ is an integer, for the same number of users $K$ and memory ratio $\frac{M}{N}$ in Corollary \ref{cor:dpda_new}, the JCM scheme gives a transmission load $R=\frac{(q-1)^t}{q^t-(q-1)^t}$ and subpacketization level $F=\frac{KM}{N}\binom{K}{\frac{KM}{N}}$.
Since $\frac{M}{N}=1-\left(\frac{q-1}{q}\right)^t$, for a fixed $q$ and $t$, $\frac{M}{N}$ is fixed. Since $K=q^{m-1}$, as $m \longrightarrow \infty$, $K \longrightarrow \infty$. Therefore, for a fixed $\frac{M}{N}$, as $K \longrightarrow \infty$, the subpacketization level of the proposed scheme in Corollary \ref{cor:dpda_new} can be approximated as follows.
\begin{equation*}
	F=\binom{m}{t}q^t \sim \binom{\frac{\log K}{\log q}}{t}q^t \sim \frac{\left(\frac{\log K}{\log q}\right)^t q^t}{t!}.
\end{equation*}
That is, the subpacketization level of the proposed scheme in Corollary \ref{cor:dpda_new} is of order $O\left({(\log K)}^t\right)$ as $K \longrightarrow \infty$. Thus, compared to the JCM scheme, the proposed scheme in Corollary \ref{cor:dpda_new} has significantly lesser subpacketization level at the expense of an increase in transmission load. 

\hspace{1em} $\bullet$ \textit{Comparison of the scheme in Corollary \ref{cor:dpda_new} with scheme obtainable by transforming regular PDA}: For any positive integers $q$, $m$ and $t$ with $q \geq 2$ and $t <m$, \cite{MACC_des} and \cite{MJXQ} give a $\binom{m}{t}-\left(\binom{m}{t}q^t, q^{m-1}, q^{m-1}-q^{m-t-1}(q-1)^t, q^{m-1}(q-1)^t \right)$ PDA. Since the number of $\star$ entries in all the rows of this PDA are same, the transpose of this PDA results in a $\binom{m}{t}-\left(q^{m-1}, \binom{m}{t}q^t, \left(q^t-(q-1)^t\right)\binom{m}{t}, (q-1)^tq^{m-1}\right)$ PDA. From this PDA, using the transformation method given in \cite{JMQX}, one can obtain a DPDA which gives a $(q^{m-1},M,N)$ D2D coded caching scheme with $F=\left[\binom{m}{t}-1\right]\binom{m}{t}q^t$, $\frac{M}{N}=1-\left(\frac{q-1}{q}\right)^t$  and $R= \frac{(q-1)^tq^{m-t-1}}{\binom{m}{t}-1}$. Compared to this scheme, for the same $K$ and $\frac{M}{N}$, the proposed scheme in Corollary \ref{cor:dpda_new} has an advantage in both subpacketization level and transmission load. In fact, the DPDA in Claim \ref{claim:dpda_new} is exactly the same as the transposed PDA of the $\binom{m}{t}-\left(\binom{m}{t}q^t, q^{m-1}, q^{m-1}-q^{m-t-1}(q-1)^t, q^{m-1}(q-1)^t \right)$ PDA in  \cite{MACC_des} and \cite{MJXQ}.

\hspace{1em} $\bullet$ \textit{Comparison of the proposed D2D schemes with schemes in \cite{JY}}: Four classes of DPDAs were constructed in \cite{JY}. A general comparison of the proposed D2D schemes with these schemes is not possible, since the system parameters $K$ and $\frac{M}{N}$ are not identical. A comparison is carried out in Table \ref{tab:D2D_comp} considering instances where different schemes have the same system parameters. As evident from Table \ref{tab:D2D_comp}, compared to the schemes in \cite{JY}, the proposed schemes have an advantage in subpacketization at the expense of an increase in load, or vice versa. 
 \begin{table}[!htbp]
 	\centering
 	\setlength{\tabcolsep}{3pt}
 	\scriptsize
 	\begin{tabular}{|c|c| c | c | c | c |}
 		\hline
 		\rule{0pt}{4ex}
 		\makecell{Sl. No.}&\makecell{Schemes and Parameters}  & \makecell{$K$} & \makecell{$\frac{M}{N}$} & \makecell{$F$}& \makecell{$R$} \\ [4pt] 	
 		\hline
 		\rule{0pt}{3.5ex}
 		1 & \makecell{Proposed scheme in Corollary \ref{cor:dpda3} \\ for $n=11$, $k=5$ and $j=2$} &
 		\multirow{2}{*}{$462$} &
 		\multirow{2}{*}{$\frac{9}{11}$} &
 		$825$ & $14$ \\
 		\cline{1-2}\cline{5-6}
 		\rule{0pt}{3.5ex}
 		2 & \makecell{Theorem $5$ in \cite{JY} \\ for $n=11$, $a=5$ and $b=7$} &
 		& & $330$ & $28$ \\
 		\hline
 		\rule{0pt}{3.5ex}
 		3 & \makecell{Proposed scheme in Corollary \ref{cor:dpda4} \\ for $m=7$, $t=2$ and $q=2$} &
 		\multirow{2}{*}{$84$} &
 		\multirow{2}{*}{$\frac{3}{4}$} &
 		$128$ & $3$ \\
 		\cline{1-2}\cline{5-6}
 		\rule{0pt}{3.5ex}
 		4 & \makecell{Theorem $2$ in \cite{JY} \\ for $n=9$, $a=3$ and $t=1$} &
 		& & $72$ & $1.167$ \\
 		\hline
 		\rule{0pt}{3.5ex}
 		5 & \makecell{Proposed scheme in Corollary \ref{cor:dpda3} \\ for $n=5$, $k=2$ and $j=1$} &
 		\multirow{2}{*}{$10$} &
 		\multirow{2}{*}{$\frac{3}{5}$} &
 		$5$ & $4$ \\
 		\cline{1-2}\cline{5-6}
 		\rule{0pt}{3.5ex}
 		6 & \makecell{Theorem $3$ in \cite{JY} \\ for $n=5$ and $a=1$} &
 		& & $10$ & $1$ \\
 		\hline
 		\rule{0pt}{3.5ex}
 		7 & \makecell{Proposed scheme in Corollary \ref{cor:dpda3} \\ for $n=17$, $k=3$ and $j=1$} &
 		\multirow{2}{*}{$680$} &
 		\multirow{2}{*}{$\frac{14}{17}$} &
 		$238$ & $60$ \\
 		\cline{1-2}\cline{5-6}
 		\rule{0pt}{3.5ex}
 		8 & \makecell{Theorem $5$ in \cite{JY} \\ for $n=17$, $a=3$ and $b=5$} &
 		& & $6188$ & $\frac{20}{13}$ \\
 		\hline
 	\end{tabular}
 	\caption{Comparison of the proposed D2D schemes with the schemes in \cite{JY}}
 	\label{tab:D2D_comp}
 \end{table}
\section{Conclusion and Directions for future work}\label{concl_madcc}
In this paper, we considered an MADCC network with access topologies derived from combinatorial $t$-designs and $t$-GDDs, and developed novel schemes. Using the metrics of load per user and subpacketization level, we compared the proposed schemes with the derived CWEC schemes and the WCCWC scheme while keeping the number of caches, cache size, and access degree the same. Further, we obtained several novel classes of low subpacketization level coded caching schemes for the original D2D network, where each user has its own cache. The only MADCC schemes studied so far in the literature are the WCCWC scheme with cyclic wrap-around topology and the proposed schemes with $t$-design and $t$-GDD access topologies. We list some practical limitations of these schemes and outline directions for future research.

$\bullet$ \textit{Decentralized caching:} All known MADCC schemes, including the proposed ones, rely on centralized deterministic cache placement, which requires prior knowledge of the users participating in the delivery phase and the access topology. While this enables multicast coding gains, real D2D wireless networks are dynamic; users may join or leave the network over time, and cache nodes can turn on or off. Therefore, a decentralized random caching scheme, in which cache placement is performed independently and without global coordination, is more suitable for practical scenarios such as fully decentralized IoT networks. Unlike shared-link networks, where the server can always supply file parts missing due to decentralized placement, D2D networks cannot recover uncached portions of files. Hence, a properly chosen Maximum Distance Separable (MDS) encoding must be applied before random caching to guarantee successful decoding. The decentralized caching scheme for the original D2D network in \cite{Ji, TS} adopts this approach. Thus, decentralized MADCC schemes with $t$-design and $t$-GDD access topologies could be developed by first applying an appropriate MDS encoding, followed by decentralized random caching, and then leveraging the structured overlap patterns created by the uniform block-intersection properties of these combinatorial designs to obtain coded multicast opportunities. This is an interesting direction for future research.  

$\bullet$ \textit{Generalized access topologies:} A multiaccess D2D model places no structural constraints on user access sets; each user may access any arbitrarily chosen subset of caches. Existing MADCC schemes have the limitation that the access topology is fixed and cannot accommodate changes in the access topology. Developing MADCC schemes for other possible access topologies, as well as designing placement strategies that are agnostic to the access topology, are important directions for future research. 

$\bullet$ \textit{Non-uniform file popularity:}  Like other placement and delivery arrays based coded caching schemes (shared-link or D2D, with dedicated cache or multiaccess), the proposed MADCC schemes with combinatorial access topologies also give equal consideration to all files during cache placement, regardless of their popularity. However, in practical scenarios, file popularity typically follows arbitrary distributions, and memory allocation should therefore account for these non-uniform popularity profiles. To the best of our knowledge, the only existing work that studied the original D2D coded caching network with non-uniform file popularity is \cite{LZZ}, which proposed an achievable scheme by classifying files into two groups: popular and non-popular files. In \cite{HKD}, the authors showed that for a multiaccess shared-link coded caching network with non-uniform popularity, a multi-level popularity classification approach with more than two groups is optimal within a constant factor in load. Therefore, studying optimization or learning-based methods to optimize the performance metrics under consideration in MADCC networks with non-uniform file popularity is a relevant research direction.

$\bullet$ \textit{Selfish user behaviors:} Apart from dynamic user arrivals or departures, practical D2D communication scenarios often involve users exhibiting selfish behavior \cite{GZCLJC}. A user who does not participate in transmitting data during the delivery phase is termed a selfish user. Such behavior may arise from the desire to conserve energy or from privacy and security concerns \cite{GZCLJC}. Despite not transmitting, selfish users may still request files. Coded caching in an original D2D network consisting of selfish users (referred to as a \textit{partially cooperative D2D network}) was first studied by Tebbi and Sung in \cite{TS} and subsequently in other works. In the proposed MADCC schemes with $t$-design and $t$-GDD topologies, exactly $\lambda$ and $\binom{m}{t}$ users, respectively, can perform the coded multicast transmission corresponding to any non-$\star$ entry in the delivery array. Consequently, schemes derived from $t$-designs with $\lambda > 1$ can tolerate up to $\lambda-1$ selfish users without any performance degradation, and schemes derived from and $t$-GDDs can tolerate up to $\binom{m}{t}-1$ selfish users while maintaining the same performance. However, designing MADCC schemes specifically for accommodating any number of selfish users is a good direction of research. 
  
 Developing MADCC schemes that jointly address more than one of the above-mentioned limitations of existing MADCC schemes is of great significance.
 \section*{Acknowledgement}
 This work was supported partly by the Science and Engineering Research Board (SERB) of Department of Science and Technology (DST), Government of India, through J.C Bose National Fellowship to B. Sundar Rajan.	
\begin{appendices}
	\section{Proof of Claim \ref{claim:dpda_tdes}}\label{appendix:dpda_tdes}
	By (\ref{eq:da_tdes}), a non-$\star$ entry $\left(\mathcal{D} \cup A(\mathcal{T})\right)_{\alpha(\mathcal{D},A(\mathcal{T}))}$ occur only when $A \cap \mathcal{D} = \emptyset$. Therefore, $A(\mathcal{T}) \cap \mathcal{D} = \emptyset$ and  $|\mathcal{D} \cup A(\mathcal{T})|=i+(t-i)=t$. Therefore, the number of possible $\mathcal{D} \cup A(\mathcal{T})$ is $\binom{v}{t}$. Consider a specific non-$\star$ entry $\left(\mathcal{D} \cup A(\mathcal{T})\right)_{\alpha(\mathcal{D},A(\mathcal{T}))}$. Then, by Theorem \ref{thm:tdes2}, there are exactly $\lambda_{t-i}^{i}=\frac{\lambda \binom{v-(t-i)-i}{k-(t-i)}}{\binom{v-t}{k-t}}=\frac{\lambda \binom{v-t}{k-t+i}}{\binom{v-t}{k-t}}$ blocks in $\mathcal{A}$ that contains all the points in $A(\mathcal{T})$ and none of the points in $\mathcal{D}$. Therefore by (\ref{eq:da_tdes}), for a given $\mathcal{D}$, a specific $\mathcal{D} \cup A(\mathcal{T})$ occurs  $\frac{\lambda \binom{v-t}{k-t+i}}{\binom{v-t}{k-t}}$ times in $\mathbf{Q}$.   Therefore, $\alpha(\mathcal{D},A(\mathcal{T})) \in \left[\frac{\lambda \binom{v-t}{k-t+i}}{\binom{v-t}{k-t}}\right]$. Thus, the number of non-$\star$ entries $S=\binom{v}{t}\frac{\lambda \binom{v-t}{k-t+i}}{\binom{v-t}{k-t}}$.

	From (\ref{eq:da_tdes}), it can be seen that the non-$\star$ entries in each column $A \in \mathcal{A}$ occur in those rows $(\mathcal{D},\mathcal{T})$ such that $A \cap \mathcal{D} = \emptyset$. Since $|\mathcal{D}|=i$ and $|A|=k$, there are $\binom{v-k}{i}$ non-$\star$ entries, for a given $\mathcal{T}$, in each column. Therefore, each column has $\binom{v-k}{i}\binom{k}{t-i}$ non-$\star$ entries. That is, the number of $\star$s in each column of $\mathbf{Q}$ is $Z=\left(\binom{v}{i}-\binom{v-k}{i}\right)\binom{k}{t-i}$. Thus $C1$ of DPDA definition holds.  $C2$ is obvious by construction. 
	
	Consider the non-$\star$ entries $\left(\mathcal{D} \cup A(\mathcal{T})\right)_{\alpha(\mathcal{D},A(\mathcal{T}))}$ appearing in a column $A$.  Since $A(\mathcal{T})$ represents the points in $A$ with index in $\mathcal{T}$, $A(\mathcal{T})$ is different for different $\mathcal{T}$. For a given $\mathcal{T}$, the $\mathcal{D}$ in the row indices $(\mathcal{D},\mathcal{T})$ are different.  Thus $\left(\mathcal{D} \cup A(\mathcal{T})\right)_{\alpha(\mathcal{D},A(\mathcal{T}))}$ appears only once in a given column $A$. Since $\alpha(\mathcal{D},A(\mathcal{T}))$ denotes the occurrence number of $\mathcal{D} \cup A(\mathcal{T})$ in $\mathbf{Q}$ from top to bottom and left to right, $\left(\mathcal{D} \cup A(\mathcal{T})\right)_{\alpha(\mathcal{D},A(\mathcal{T}))}$ cannot appear more than once in a row. Thus $C3.(a)$ of DPDA definition holds. 
	
	For any two distinct entries $Q_{(\mathcal{D}_i,\mathcal{T}_i),A_i}$, $Q_{(\mathcal{D}_j,\mathcal{T}_j),A_j}$ with $(\mathcal{D}_i,\mathcal{T}_i) \neq (\mathcal{D}_j,\mathcal{T}_j)$ and $A_i \neq A_j$, let $ Q_{(\mathcal{D}_i,\mathcal{T}_i),A_i} = Q_{(\mathcal{D}_j,\mathcal{T}_j),A_j}$. Then by (\ref{eq:da_tdes}), $\mathcal{D}_i \cup A_i(\mathcal{T}_i)= \mathcal{D}_j \cup A_j(\mathcal{T}_j)$, $\mathcal{D}_i \cap A_i(\mathcal{T}_i)= \mathcal{D}_j \cap A_j(\mathcal{T}_j)=\emptyset$ and $|\mathcal{D}_i|=|\mathcal{D}_j|$. Therefore, $\mathcal{D}_i \cap A_j(\mathcal{T}_j) \ne \emptyset$ and $\mathcal{D}_j \cap A_i(\mathcal{T}_i) \ne \emptyset$. That is $\mathcal{D}_i \cap A_j \ne \emptyset$ and $\mathcal{D}_j \cap A_i \ne \emptyset$. Therefore, by (\ref{eq:da_tdes}), $Q_{(\mathcal{D}_i,\mathcal{T}_i),A_j}= \star $ and $Q_{(\mathcal{D}_j,\mathcal{T}_j),A_i}= \star $. Thus $C3.(b)$ of DPDA definition holds.
	
	We define the mapping $\phi$ from the non-$\star$ entries $\left(\mathcal{D} \cup A(\mathcal{T})\right)_{\alpha(\mathcal{D},A(\mathcal{T}))}$  to the users $\{U_A : A \in \mathcal{A}\}$ as $\{\phi\left(\left(\mathcal{D} \cup A(\mathcal{T})\right)_{\alpha(\mathcal{D},A(\mathcal{T}))}\right)= U_{A_i} : \mathcal{D} \cup A(\mathcal{T}) \subset A_i \} $. Since $|\mathcal{D} \cup A(\mathcal{T})|=t$, there are exactly $\lambda$ users satisfying this condition. Also, since $\mathcal{D} \cup A(\mathcal{T}) \subset A_i$, by (\ref{eq:da_tdes}), $Q_{(\mathcal{D},\mathcal{T}),A_i}= \star $. Thus $C4$ of DPDA definition holds. \hfill $\blacksquare$.
	
	\section{Proof of Corollary \ref{cor:MADCC_tdes}}\label{appendix:dpda_tdes_cor}
		The MADCC scheme in Corollary \ref{cor:MADCC_tdes} is obtained by slightly modifying the  MADCC scheme in Theorem \ref{thm:MADCC_tdes} as follows. The central server first splits each file $\{W_n : n \in [N]\}$ into $\binom{v}{i}$, where $i \in [\min(t,k-t+1)]$, non-overlapping packets and then further divide each packet $W_{n,\mathcal{D}}$ into  $\binom{k}{t-1}$ sub-packets, i.e., $\{W_{n,\mathcal{D}}=W_{n,\mathcal{D}}^{\mathcal{T}} : \mathcal{T} \in \binom{[k]}{t-1}, \mathcal{D} \in \binom{[v]}{i}, n \in [N]\}$. Therefore, $F=\binom{v}{i}\binom{k}{t-1}$. The  $\binom{v}{i}\binom{k}{t-1} \times v$ placement array $\mathbf{Z}=\left(Z_{(\mathcal{D},\mathcal{T}),x}\right)_{\mathcal{D} \in \binom{[v]}{i}, \mathcal{T} \in \binom{[k]}{t-1}, x \in \mathcal{X} }$ and the $\binom{v}{i}\binom{k}{t-1} \times \frac{\lambda \binom{v}{t}}{\binom{k}{t}}$ delivery array $\mathbf{Q}=\left(Q_{(\mathcal{D},\mathcal{T}),A} \right)_{\mathcal{D} \in \binom{[v]}{i}, \mathcal{T} \in \binom{[k]}{t-1}, A \in \mathcal{A} }$ are defined as in (\ref{eq:pa_tdes}) and  (\ref{eq:da_tdes}) respectively. 
		
		The resulting delivery array will be a $\left(\frac{\lambda \binom{v}{t}}{\binom{k}{t}}, \binom{v}{i}\binom{k}{t-1}, \left(\binom{v}{i}-\binom{v-k}{i}\right)\binom{k}{t-1},  \frac{\lambda\binom{v}{i+t-1}\binom{v-(i+t-1)}{k-t+1}}{\binom{v-t}{k-t}} \right)$ DPDA. In a $t$-design, the properties and uniformity of subsets of size greater than $t$ cannot be generalized. For a $t-(v,k,\lambda)$ design $(\mathcal{X},\mathcal{A})$, let $\mathcal{D},\mathcal{Y} \subseteq \mathcal{X}$, where $\mathcal{D} \cap \mathcal{Y} =\emptyset$, $|\mathcal{D}|=i \le t$,$|\mathcal{Y}|=t-1$, then by \textit{the principle of inclusion-exclusion}, the number of blocks $A \in \mathcal{A}$ that contain all the points in $\mathcal{Y}$ and none of the points in $\mathcal{D}$, $\lambda_{t-1}^i(\mathcal{D},\mathcal{Y}) = \lambda(\mathcal{Y}) + \sum\limits_{\substack{\mathcal{D'} \subseteq \mathcal{D} : |\mathcal{D'}| \in [j] \\ j=1}}^{i} (-1)^j \lambda\left(\mathcal{D'} \cup \mathcal{Y} \right)$. $\lambda(\mathcal{Y})=\lambda_{t-1}$ and $\lambda\left(\mathcal{D'} \cup \mathcal{Y} \right)$ may not be the same for all $\mathcal{D'}$ and $\mathcal{Y}$. Our conjecture is that, $\lambda\left(\mathcal{D'} \cup \mathcal{Y} \right)$ is same for all the $\mathcal{D'}$ and $\mathcal{Y}$ such that $\mathcal{D'} \cup \mathcal{Y} \subseteq A \in \mathcal{A}$.
		
		The $|\mathcal{D}\cup A(\mathcal{T})| = i+t-1$ and $t \le i+t-1 \le k$ for $i \in [\min(t,k-t+1)]$. Since the uniformity of $\mathcal{D}\cup A(\mathcal{T})$ is different, the number of blocks in $\mathcal{A}$ that contains all the points in $A(\mathcal{T})$ and none of the points in $\mathcal{D}$, $\lambda_{t-1}^i$, will be different for different $(t-1)$ sized subsets for a given $\mathcal{D}$. The $i+t-1$ sized set will be either a subset of blocks in the design or not a subset of any block. Thus the values $\alpha(\mathcal{D},A(\mathcal{T}))$ can take, will be different for different choices of $\mathcal{D}$ and $A(\mathcal{T})$. For a given $\mathcal{D}\cup A(\mathcal{T})$, if the values $\alpha(\mathcal{D'},A(\mathcal{T}))$ can take is same for all $\mathcal{D'} \subset \mathcal{D}\cup A(\mathcal{T})$, then the regularity $g$ of that non-$\star$ entry is $\binom{i+t-1}{i}$.  Therefore, $S=\frac{K(F-Z)}{g}=\frac{\lambda \binom{v}{t}\binom{v-k}{i}\binom{k}{t-1}}{\binom{k}{t}\binom{i+t-1}{i}}=\frac{\lambda\binom{v}{i+t-1}\binom{v-(i+t-1)}{k-t+1}}{\binom{v-t}{k-t}}$. The proof that this delivery array satisfies the conditions $C1-C3$ of DPDA definition is similar to the proof in Appendix \ref{appendix:dpda_tdes}. The mapping $\phi$ from the non-$\star$ entries $\left(\mathcal{D} \cup A(\mathcal{T})\right)_{\alpha(\mathcal{D},A(\mathcal{T}))}$  to the users $\{U_A : A \in \mathcal{A}\}$ is defined as $\{\phi\left(\left(\mathcal{D} \cup A(\mathcal{T})\right)_{\alpha(\mathcal{D},A(\mathcal{T}))}\right)= U_{A_i} : |\left(\mathcal{D} \cup A(\mathcal{T})\right) \cap A_i |\geq t \} $. Since $|\left(\mathcal{D} \cup A(\mathcal{T})\right) \cap A_i |\geq t$, $|A(\mathcal{T})|=t-1$ and $\mathcal{D} \cap A(\mathcal{T})=\emptyset$, it follows that $\mathcal{D} \cap A_i \neq \emptyset $. Therefore, by (\ref{eq:da_tdes}), $Q_{(\mathcal{D},\mathcal{T}),A_i}= \star $. Thus $C4$ of DPDA definition holds. Since $\mathbf{Q}$ is a DPDA, each user's demand can be satisfied with a transmission load $R=\frac{S}{F}= \frac{\lambda\binom{v-i}{k}}{\binom{v-t}{k-t}\binom{i+t-1}{i}}$.	\hfill $\blacksquare$.

     \section{Proof of Claim \ref{claim:dpda_tgdd}}\label{appendix:dpda_tgdd}
    By Lemma \ref{lem:oa_mds}, for the $s-(q,m,1)$ OA $(\mathcal{Y}, \mathcal{D})$, $d_{min}(\mathcal{D})=m-s+1$.  By (\ref{eq:da_tgdd}) and (\ref{eq:da_tgdd2}), for any non-$\star$ entry $\vec{e}_{\alpha(\vec{e})}=\left(e_1,e_2,...,e_m\right)_{\alpha(\vec{e})}$, where $e_n \in [q], \forall n \in [m]$, occurring in a row  $(\vec{D}_j,\mathcal{T})$, $d\left(\vec{D}_j,\vec{e}\right) = l < d_{min}(\mathcal{D})$. Therefore, $\vec{e}$ will not be any of the $q^s$ rows of $\mathcal{D}$. Therefore, there are at most $q^m-q^s$ different vectors $\vec{e}$ with length $m$. For a given $\vec{e}$ in a row corresponding to a $\vec{D}_j$, let the $l$ indices at which $\vec{D}_j$ and $\vec{e}$ differ be $\{(u'_1,u'_2,...,u'_l)\}$ and the corresponding entries be $\{(v'_1,v'_2,...,v'_l)\}$. Then, for the given  $\vec{D}_j$, $\vec{e}$ will appear in all the columns $A_i \in \mathcal{A}$ such that $d\left(\vec{D}_j(\vec{\psi}(A_i)),\vec{\varphi}(A_i)\right) = k$ and  $\{(u'_1,v'_1),(u'_2,v'_2),...,(u'_l,v'_l)\} \subset A_i$. It is hard to find the exact number of occurrence of a given $\vec{e}$ in a row corresponding to a $\vec{D}_j$. By Theorem \ref{thm:tgdd}, the number of blocks $A_i \in \mathcal{A}$ which contain $\{(u'_1,v'_1),(u'_2,v'_2),...,(u'_l,v'_l)\}$ is  $\frac{ q^{t-l} \binom{m-l}{t-l}}{\binom{k-l}{t-l}}$. Therefore, $\alpha(\vec{e})$ can be at most $\frac{q^{t-l} \binom{m-l}{t-l}}{\binom{k-l}{t-l}}$. Thus, the number of non-$\star$ entries $S$ is at most $\frac{(q^m-q^s) q^{t-l} \binom{m-l}{t-l}}{\binom{k-l}{t-l}}$.

    From (\ref{eq:da_tgdd}), it can be seen that the non-$\star$ entries in each column $A_i \in \mathcal{A}$ occur in those rows $(\vec{D}_j,\mathcal{T})$ such that $d\left(\vec{D}_j(\vec{\psi}(A_i)),\vec{\varphi}(A_i)\right) = k$. For any given $A_i \in \mathcal{A}$, by the property of orthogonal array, the number of possible $\vec{D}_j$ for which  $d\left(\vec{D}_j(\vec{\psi}(A_i)),\vec{\varphi}(A_i)\right) = k$ is $(q-1)^kq^{s-k}$. For each $\vec{D}_j$, there are $ \binom{k}{l}$ rows of $\mathbf{Q}$ indexed $(\vec{D}_j,\mathcal{T})$. Therefore, each column has $(q-1)^kq^{s-k}\binom{k}{l}$ non-$\star$ entries. That is, the number of $\star$s in each column of $\mathbf{Q}$ is $Z=\left(q^s-(q-1)^kq^{s-k}\right)\binom{k}{l}$. Thus $C1$ of DPDA definition holds.  $C2$ is obvious by construction.  
    
    Next we prove by contradiction, that a given non-$\star$ entry cannot appear more than once in a column. Assume that $\vec{e}_{\alpha(\vec{e})}$  appears twice in a column $A_i$ in the rows $(\vec{D}_j,\mathcal{T})$ and $(\vec{D}_{j'},\mathcal{T'})$. Then by (\ref{eq:da_tgdd}), $d\left(\vec{D}_j(\vec{\psi}(A_i)),\vec{\varphi}(A_i)\right) = d\left(\vec{D}_{j'}(\vec{\psi}(A_i)),\vec{\varphi}(A_i)\right) = k$. By  (\ref{eq:da_tgdd2}), $d\left(\vec{D}_j,\vec{e}\right)=d\left(\vec{D}_{j'},\vec{e}\right)=l$, $\vec{e}(\vec{\psi}(A_i))(\mathcal{T})= \vec{\varphi}(A_i)(\mathcal{T})$ and $\vec{e}(\vec{\psi}(A_i))(\mathcal{T'})= \vec{\varphi}(A_i)(\mathcal{T'})$. That is, $\vec{e}(\vec{\psi}(A_i))(\mathcal{T} \cup \mathcal{T'})= \vec{\varphi}(A_i)(\mathcal{T} \cup \mathcal{T'})$. If $\mathcal{T} = \mathcal{T'}$, then we get $d\left(\vec{D}_j,\vec{D}_{j'} \right)= l < d_{min}(\mathcal{D})$, which is a contradiction. If $\mathcal{T} \ne \mathcal{T'}$, then $|\mathcal{T} \cup \mathcal{T'}| > l$. Therefore, $\vec{e}(\vec{\psi}(A_i))(\mathcal{T} \cup \mathcal{T'})= \vec{\varphi}(A_i)(\mathcal{T} \cup \mathcal{T'})$ implies either $d\left(\vec{D}_j(\vec{\psi}(A_i)),\vec{\varphi}(A_i)\right) < k$ or $d\left(\vec{D}_{j},\vec{e}\right) >l$, both of  which are contradictions. Thus $\vec{e}_{\alpha(\vec{e})}$ appear only once in a given column $A_i$. Since $\alpha(\vec{e})$ denotes the occurrence number of $\vec{e}$ in $\mathbf{Q}$ from top to bottom and left to right for a given $\vec{D}_j$, $\vec{e}_{\alpha(\vec{e})}$ cannot appear more than once in a row $(\vec{D}_j,\mathcal{T})$. Thus $C3.(a)$ of DPDA definition holds. 
    
    Consider two non-$\star$ entries $Q_{(\vec{D}_j,\mathcal{T}),A_i}$, $Q_{(\vec{D}_{j'},\mathcal{T'}),A_{i'}}$ with $(\vec{D}_j,\mathcal{T}) \neq (\vec{D}_{j'},\mathcal{T'})$ and $A_i \neq A_{i'}$, such that $ Q_{(\vec{D}_j,\mathcal{T}),A_i} = Q_{(\vec{D}_{j'},\mathcal{T'}),A_{i'}}=\vec{e}_{\alpha(\vec{e})}$. Then by (\ref{eq:da_tgdd}) and (\ref{eq:da_tgdd2}) we have, $d\left(\vec{D}_j(\vec{\psi}(A_i)),\vec{\varphi}(A_i)\right) = d\left(\vec{D}_{j'}(\vec{\psi}(A_{i'})),\vec{\varphi}(A_{i'})\right) =  k$, $d\left(\vec{D}_j,\vec{e}\right)=d\left(\vec{D}_{j'},\vec{e}\right)=l$, $\vec{e}(\vec{\psi}(A_i))(\mathcal{T})= \vec{\varphi}(A_i)(\mathcal{T})$ and $\vec{e}(\vec{\psi}(A_{i'}))(\mathcal{T'})= \vec{\varphi}(A_{i'})(\mathcal{T'})$. If   $\left(\vec{\psi}(A_i)(\mathcal{T}),\vec{\varphi}(A_i)(\mathcal{T})\right) =  \left(\vec{\psi}(A_{i'})(\mathcal{T'}), \vec{\varphi}(A_{i'})(\mathcal{T'})\right)$, then similar to the proof of condition  $C3.(a)$, we get  $d\left(\vec{D}_j(\vec{\psi}(A_{i'})),\vec{\varphi}(A_{i'})\right) < k$ and $d\left(\vec{D}_{j'}(\vec{\psi}(A_{i})),\vec{\varphi}(A_{i})\right) < k$. Then by (\ref{eq:da_tgdd}),   $Q_{(\vec{D}_j,\mathcal{T}),A_{i'}}=\star$ and $Q_{(\vec{D}_{j'},\mathcal{T'}),A_i}=\star$. If $\vec{\psi}(A_i)(\mathcal{T})=\vec{\psi}(A_{i'})(\mathcal{T'})$ and $\vec{\varphi}(A_i)(\mathcal{T})\ne \vec{\varphi}(A_{i'})(\mathcal{T'})$, then by  (\ref{eq:da_tgdd2}), $\vec{e}$ cannot be the same at $Q_{(\vec{D}_j,\mathcal{T}),A_i}$ and $Q_{(\vec{D}_{j'},\mathcal{T'}),A_{i'}}$. Now consider the case when  $\vec{\psi}(A_i)(\mathcal{T}) \ne \vec{\psi}(A_{i'})(\mathcal{T'})$.  Since  $\vec{\psi}(A_i)(\mathcal{T}) \ne \vec{\psi}(A_{i'})(\mathcal{T'})$, there exist some $u_{l}$ and  $u_{l'}$, where $l, l' \in [k]$, such that $u_{l} \in \vec{\psi}(A_i)(\mathcal{T})\backslash \vec{\psi}(A_{i'})(\mathcal{T'})$ and $u_{l'} \in \vec{\psi}(A_{i'})(\mathcal{T'})\backslash \vec{\psi}(A_i)(\mathcal{T})$. Then, by (\ref{eq:da_tgdd2}),  $e_{u_{l}}=v_{l}=\vec{D}_{j'}(u_{l})$ and $e_{u_{l'}}=v_{l'}=\vec{D}_j(u_{l'})$. Since $v_{l}=\vec{D}_{j'}(u_{l})$ and $v_{l'}=\vec{D}_j(u_{l'})$, $d\left(\vec{D}_{j'}(\vec{\psi}(A_{i})),\vec{\varphi}(A_{i})\right) < k$ and $d\left(\vec{D}_j(\vec{\psi}(A_{i'})),\vec{\varphi}(A_{i'})\right) < k$. Then by (\ref{eq:da_tgdd}),  $Q_{(\vec{D}_{j'},\mathcal{T'}),A_i}=\star$ and $Q_{(\vec{D}_j,\mathcal{T}),A_{i'}}=\star$. Thus $C3.(b)$ of DPDA definition holds.                                                                                                           
    
    We define the mapping $\phi$ from the non-$\star$ entries $\vec{e}_{\alpha(\vec{e})}$  to the users $\{U_{A_{i'}} : A_{i'} \in \mathcal{A}\}$ as $\{\phi\left(\vec{e}_{\alpha(\vec{e})}\right)= U_{A_{i'}} : \vec{e}(\vec{\psi}(A_{i'})(\mathcal{Y})) \subseteq \vec{\varphi}(A_{i'}), \text{ for some } \mathcal{Y} \in \binom{[k]}{t} \}$. In other words, for $\vec{e}=\left(e_1,e_2,...,e_m\right)$, $\phi\left(\vec{e}_{\alpha(\vec{e})}\right)$ are those blocks $U_{A_{i'}}$ which contains any of the $t$-set of points among the points $\{(1,e_1),(2,e_2),...,(m,e_m)\}$. By definition of $t$-GDD, every $t$-set of points from $t$ distinct groups is in exactly $\lambda=1$ block. Therefore, for any non-$\star$ entry, there exist users with the above mapping, and at most there will be $\binom{m}{t}$ such users.  When $k=t$, for each non-$\star$ entry, there are exactly $\binom{m}{t}$ users satisfying this condition.  If $\vec{e}(\vec{\psi}(A_{i'})(\mathcal{Y})) \subseteq \vec{\varphi}(A_{i'})$, then $d\left(\vec{e}(\vec{\psi}(A_{i'})),\vec{\varphi}(A_{i'})\right) \le k-t$.  Let $Q_{(\vec{D}_{j},\mathcal{T}),A_{i}}= \vec{e}_{\alpha(\vec{e})}$. Then by (\ref{eq:da_tgdd2}), $d\left(\vec{D}_{j},\vec{e}\right)=d\left(\vec{D}_{j}(\vec{\psi}(A_i))(\mathcal{T})  ,\vec{e}(\vec{\psi}(A_i))(\mathcal{T})\right)=l$. Thus,  $d\left(\vec{D}_j(\vec{\psi}(A_{i'})), \vec{\varphi}(A_{i'})\right) \le d\left(\vec{D}_j(\vec{\psi}(A_{i'})),\vec{e}(\vec{\psi}(A_{i'}))\right) + d\left(\vec{e}(\vec{\psi}(A_{i'})),\vec{\varphi}(A_{i'})\right) \le d\left(\vec{D}_{j},\vec{e}\right) + k-t = l +k-t <   k$, since $l<t$. Therefore by (\ref{eq:da_tgdd}),  $Q_{(\vec{D}_j,\mathcal{T}),A_{i'}}=\star$. Thus $C4$ of DPDA definition holds.  
 Therefore, $\mathbf{Q}$ is a $\left(\frac{ \binom{m}{t}q^t}{\binom{k}{t}},  q^s\binom{k}{l},  \left( q^s- (q-1)^k q^{s-k}\right)\binom{k}{l}, S \right)$ DPDA, where $S \le {(q^m-q^s) q^{t-l} \binom{m-l}{t-l}}/{\binom{k-l}{t-l}}$. 	\hfill $\blacksquare$. 

\section{Computation of exact value of $S$ in Claim \ref{claim:dpda_tgdd}}\label{appendix:S_tgdd}
The values $\alpha(\vec{e})$ takes can be exactly found when $k=t$. When $k=t$, the columns $A_i \in \mathcal{A}$ such that $d\left(\vec{D}_j(\vec{\psi}(A_i)),\vec{\varphi}(A_i)\right) = k$ is noting but the blocks of a $t-(m,q-1,t,1)$ GDD. Therefore, the number of non-star entries in a given row is $(q-1)^t \binom{m}{t}$. In this, the blocks $A_i$ such that $\{ (u'_1,v'_1),(u'_2,v'_2),...,(u'_l,v'_l) \} \subset A_i$, by Theorem \ref{thm:tgdd}, is $(q-1)^{t-l} \binom{m-l}{t-l}$. Therefore, $\alpha(\vec{e}) \in \left[ (q-1)^{t-l} \binom{m-l}{t-l} \right]$. By Lemma\ref{lem:oa_mds} and (\ref{eq:da_tgdd2}), $d\left(\vec{D}_j,\vec{e}\right) = l < d_{min}(\mathbf{D})$. Therefore, $\vec{e}$ will not be any of the $q^s$ rows of $\mathbf{D}$. 
When $k=t$ and $s=m-1$, for $l=1$, it is obvious that for any $\vec{e}$, there exist a $\vec{D}_j$ such that $d\left(\vec{D}_j,\vec{e}\right) = 1$. Therefore, the number of distinct vectors $\vec{e}$ in $\mathbf{Q}$ is equal to $q^m-q^{m-1}$. Thus, $S=(q^m-q^{m-1})(q-1)^{t-1} \binom{m-1}{t-1}=q^{m-1}(q-1)^{t} \binom{m-1}{t-1}$.

Now for $k=t$ and $s=m-t+1$ with $s >t$, we find the exact number of possible vectors $\vec{e}$ in $\mathbf{Q}$, using the proof technique used in \cite{MJXQ}. When $s=m-t+1$, i.e. $m-s=t-1$, consider $l=m-s$.  A vector $\vec{e}$ appears in $\mathbf{Q}$ if and only if there exist a row  $\vec{D}_j$ such that $d\left(\vec{D}_j,\vec{e}\right) = l=m-s$. By Theorem \ref{thm:oa_mds}, the $m-t+1-(q,m,1)$ OA is equivalent to a $[m,m-t+1]_q$ MDS code. Therefore, by Lemma \ref{lem:mds}, we have $\bigcup_{{\vec{D}_j}\in \mathcal{D}}\mathcal{E}_{\mathcal{D}}({\vec{D}_j},t-1)=\bigcup_{{\vec{D}_j}\in \mathcal{D}}\{{\vec{e}} \in\mathbb{F}^m_q \ |\ d({\vec{e}},{\vec{D}_j})\leq t-1\}=\mathbb{F}^m_q$. Therefore, for any vector $\vec{e}$, there exist a $\vec{D}_j$ satisfying $0 < d({\vec{e}},{\vec{D}_j}) \le t-1$. If $d({\vec{e}},{\vec{D}_j}) = t-1 =l$, $\vec{e}$ appears in $\mathbf{Q}$. If $0 < d({\vec{e}},{\vec{D}_j}) < t-1$,since $s>t$, then there exist a vector $\vec{e'}$ satisfying (i) $d({\vec{e'}},{\vec{D}_j}) = t-1$ and (ii) $\vec{e}$ is located on the line generated by $\vec{D}_j$ and $\vec{e'}$. That is, $\vec{e}=\alpha\vec{D}_j+\beta\vec{e'}$ for some $\alpha, \beta \in \mathbb{F}_q$. Therefore, $\vec{e}=(\alpha+\beta)\vec{D}_j+\beta(\vec{e'}-\vec{D}_j)$ and that implies, $d({\vec{e}},(\alpha+\beta){\vec{D}_j}) = d(\beta{\vec{e'}},\beta{\vec{D}_j}) = t-1$. Since $(\alpha+\beta){\vec{D}_j} \in  \mathcal{D}$, $\vec{e}$ appears in $\mathbf{Q}$. Therefore, the number of distinct vectors $\vec{e}$ in $\mathbf{Q}$ is equal to $q^m-q^{m-t+1}$. For $k=t$ and $l=t-1$, $\alpha(\vec{e}) \in \left[ (q-1) (m-t+1) \right]$. Therefore, $S=(q^m-q^{m-t+1})(q-1) (m-t+1)$. 
\hfill $\blacksquare$.

\section{Proof of Corollary \ref{cor:dpda3}}\label{appendix:dpda3}
For positive integers $t$, $m$ and $n$ such that $t \le m < n$, taking every $m$-subset of an $n$-set results in a $t-\left(n,m,\binom{n-t}{m-t}\right)$ design. Therefore, for a positive integer $k \le \frac{n}{2}$, $m=k$ and $t=k-1$, a $(k-1)-(n,k,n-k+1)$ design is obtained. For this design, Corollary \ref{cor:dpda1} gives an $\left(\binom{n}{k}, \binom{n}{i}\binom{k}{i+1}, \left(\binom{n}{i}-\binom{n-k}{i}\right)\binom{k}{i+1}, \binom{n}{k-1}\binom{n-k+1}{i+1} \right)$ DPDA, for $i \in [k-2]$. This DPDA results in an $(\binom{n}{k},M,N)$ D2D coded caching scheme with $\frac{M}{N}=1-\frac{\binom{n-k}{i}}{\binom{n}{i}}$ and $R=\frac{\binom{n-i}{k}}{\binom{k-1}{i}}$. For $k \le \frac{n}{2}$, $m=n-k$ and $t=k-1$, a $(k-1)-\\(n,n-k,\binom{n-k+1}{k})$ design is obtained. For this design, Corollary \ref{cor:dpda1} gives a $\left(\binom{n}{k}, \binom{n}{j}\binom{n-k}{k-j-1},  \left(\binom{n}{j}-\binom{k}{j}\right)\binom{n-k}{k-j-1}, \binom{n}{k-1}\right. \\ \left.\binom{n-k+1}{k-j} \right)$ DPDA, for $j \in [\min(n-k-2,k-1)]$. This DPDA results in an $(\binom{n}{k},M,N)$ D2D coded caching scheme with  $\frac{M}{N}=1-\frac{\binom{k}{j}}{\binom{n}{j}}$ and $R=\frac{\binom{n-j}{k-j}}{\binom{k-1}{j}}$. 	\hfill $\blacksquare$.

\section{Proof of Claim \ref{claim:dpda_new}}\label{appendix:dpda_new}
The number of users $K$ is equal to the number of columns in the array $\mathbf{P}$ and therefore, $K=q^{m-1}$. The subpacketization level $F$ is equal to the number of rows in the array  $\mathbf{P}$ and therefore, $F=\binom{m}{t}q^t$. In \cite{MJXQ}, a $q^{m-1} \times \binom{m}{t}q^t$ PDA is constructed using the same construction rules (with different notations) as those in (\ref{eq:d2d_new}) and (\ref{eq:d2d_new2}), except that the row and column indices are interchanged. The resulting PDA is a $\binom{m}{t}-\left(\binom{m}{t}q^t, q^{m-1}, q^{m-1}-q^{m-t-1}(q-1)^t,  q^{m-1}(q-1)^t \right)$ PDA. Since the number of $\star$ entries in all the rows of this PDA are same, the transpose of this PDA results in a $\binom{m}{t}-\left(q^{m-1}, \binom{m}{t}q^t, \left(q^t-(q-1)^t\right)\binom{m}{t}, (q-1)^tq^{m-1}\right)$ PDA. Thus, the array $\mathbf{P}$ is a PDA.  

We now show that the PDA $\mathbf{P}$ satisfies the additional condition $C4$ in the DPDA definition, thus proving it to be a DPDA. We define the mapping $\phi$ from the non-$\star$ entries $\vec{e}_{\alpha(\vec{e})}$  to the users denoted by the rows of $\mathcal{D}$, i.e., $\{\vec{D}_j, j \in [q^{m-1}]\}$ as $\{\phi\left(\vec{e}_{\alpha(\vec{e})}\right)= \vec{D}_{j} : d\left(\vec{D}_{j},\vec{e}\right) \le 1 \}$. By Theorem \ref{thm:oa_mds}, the $m-1-(q,m,1)$ OA is equivalent to a $(m,q^{m-1},2,q)$ MDS code $\mathcal{C}$, where the codewords in $\mathcal{C}$ are the rows of $\mathbf{D}$. That is $\{\vec{D}_j \in \mathcal{C}, j \in [q^{m-1}]\}$. Consider a non-$\star$ entry $\vec{e}_{\alpha(\vec{e})}$. If $\vec{e} \in \mathcal{C}$, then there exist a row $\vec{D}_{j} = \vec{e}$ and therefore $d\left(\vec{D}_{j},\vec{e}\right) =0$. If $\vec{e} \notin \mathcal{C}$, then for every $\vec{e} \notin \mathcal{C}$, there exist at least one codeword $\vec{D}_{j} \in  \mathcal{C}$ such that $d\left(\vec{D}_{j},\vec{e}\right) =1$, since $\mathcal{C}$ is a $(m,q^{m-1},2,q)$ MDS code. Therefore, corresponding to every non-$\star$ entry there exist a user mapping. Consider the row in which $\vec{e}_{\alpha(\vec{e})}$ appear to be $A_i$. Then by (\ref{eq:d2d_new2}), $\vec{e}(\vec{\psi}(A_{i}))=\vec{\varphi}(A_{i})$. Therefore, $d\left(\vec{D}_j(\vec{\psi}(A_i)),\vec{\varphi}(A_i)\right) = d\left(\vec{D}_j(\vec{\psi}(A_i)),\vec{e}(\vec{\psi}(A_{i}))\right) \le d\left(\vec{D}_{j},\vec{e}\right) \le 1 <t$. Since, $d\left(\vec{D}_j(\vec{\psi}(A_i)),\vec{\varphi}(A_i)\right) < t$, by (\ref{eq:d2d_new}), $P_{A_i,\vec{D}_{j}}=\star$. Thus $C4$ of DPDA definition holds.  Therefore, $\mathbf{P}$ is a $\left(q^{m-1}, \binom{m}{t}q^t, \left(q^t-(q-1)^t\right)\binom{m}{t}, (q-1)^tq^{m-1}\right)$ DPDA.  \hfill $\blacksquare$.
		
	\end{appendices}

\end{document}